\makeatletter
\@ifundefined{@parse@version@dash}{%
\def\@parse@version#1{\@parse@version@0#1}
\def\@parse@version@#1/#2/#3#4#5\@nil{%
\@parse@version@dash#1-#2-#3#4\@nil}
\def\@parse@version@dash#1-#2-#3#4#5\@nil{%
  \if\relax#2\relax\else#1\fi#2#3#4 }
}{} 
\makeatother

\documentclass[%
 reprint,
superscriptaddress,
nofootinbib,
 amsmath,amssymb,    
 aps,
]{revtex4-2}

\usepackage{color}
\usepackage{graphicx}
\usepackage{dcolumn}
\usepackage{bm}
\RequirePackage{siunitx}   		
\DeclareSIUnit\eVperc{\eV\per\clight}
\DeclareSIUnit\clight{\text{\ensuremath{c}}}
\RequirePackage{booktabs}  
\RequirePackage{multirow}		
\RequirePackage{pgf}
\RequirePackage{hyperref}

\newcommand{\diff}[3][]{\frac{\partial^{#1}{#2}}{\partial{#3}^{#1}}}
\newcommand{\intdif}[3]{\int_{#1}^{#2}\!\!\!\textnormal{d}{#3}}
\newcommand{\expo}[1]{\textnormal{e}^{#1}}

\newcommand{\ii}{\textnormal{i}}
\newcolumntype{d}[1]{D{.}{.}{#1}}
\newcommand{\mc}[1]{\multicolumn{1}{c}{#1}}
\newcommand{\nuc}[2]{{}^{#1}\textnormal{#2}}
\renewcommand{\H}[1]{\nuc{1}{H}}
\newcommand{\T}{\nuc{3}{H}}
\newcommand{\He}[1]{\nuc{#1}{He}} 
\newcommand{\Li}[1]{\nuc{#1}{Li}}
\newcommand{\Be}[1]{\nuc{#1}{Be}}

\begin{document} 


\title{The electromagnetic fine-structure constant in primordial nucleosynthesis revisited}

\author{Ulf-G.~Mei{\ss}ner}
\email{meissner@hiskp.uni-bonn.de}%
\affiliation{Helmholtz-Institut~f\"{u}r~Strahlen-~und~Kernphysik,%
~Rheinische~Friedrich-Wilhelms~Universit\"{a}t~Bonn,~D-53115~Bonn,~Germany}
\affiliation{Bethe~Center~for~Theoretical~Physics,%
~Rheinische~Friedrich-Wilhelms~Universit\"{a}t~Bonn,~D-53115~Bonn,~Germany}
\affiliation{Institute~for~Advanced~Simulation~(IAS-4),%
~Forschungszentrum~J\"{u}lich,~D-52425~J\"{u}lich,~Germany}
\affiliation{Institut~f\"{u}r~Kernphysik~(IKP-3)~and~J\"ulich~Center~for~Hadron~Physics,%
~Forschungszentrum~J\"{u}lich,~D-52425~J\"{u}lich,~Germany}
\affiliation{Tbilisi State University, 0186 Tbilisi, Georgia}
\author{Bernard~Ch.~Metsch}%
\email{metsch@hiskp.uni-bonn.de}%
\affiliation{Institute~for~Advanced~Simulation~(IAS-4),%
~Forschungszentrum~J\"{u}lich,~D-52425~J\"{u}lich,~Germany}
\affiliation{Helmholtz-Institut~f\"{u}r~Strahlen-~und~Kernphysik,%
~Rheinische~Friedrich-Wilhelms~Universit\"{a}t~Bonn,~D-53115~Bonn,~Germany}
\author{Helen~Meyer}%
\email{hmeyer@hiskp.uni-bonn.de}%
\affiliation{Helmholtz-Institut~f\"{u}r~Strahlen-~und~Kernphysik,%
~Rheinische~Friedrich-Wilhelms~Universit\"{a}t~Bonn,~D-53115~Bonn,~Germany}
\affiliation{Bethe~Center~for~Theoretical~Physics,%
~Rheinische~Friedrich-Wilhelms~Universit\"{a}t~Bonn,~D-53115~Bonn,~Germany}

\date{\today}

\begin{abstract}
  We study the dependence of the primordial nuclear abundances as a
  function of the electromagnetic fine-structure constant $\alpha$,
  keeping all other fundamental constants fixed. We update the leading
  nuclear reaction rates, in particular the electromagnetic contribution
  to the neutron-proton mass difference pertinent to $\beta$-decays, and
  go beyond certain approximations made in the literature. In
  particular, we include the temperature-dependence of the leading
  nuclear reactions rates and assess the systematic uncertainties by
  using four different publicly available codes for Big Bang
  nucleosynthesis.  Disregarding the unsolved so-called
  lithium-prob\-lem, we find that the current values for the
  observationally based $\nuc{2}{H}$ and $\nuc{4}{He}$ abundances
  restrict the fractional change in the fine-structure constant to less
  than $2\%$\,, which is a tighter bound than found in earlier works on
  the subject.
\end{abstract}

\maketitle


\section{\label{sec:intro}Introduction}

Since the early work of Dirac~\cite{Dirac:1973gk}, variations of the
fundamental constants of physics have been considered in a variety of
scenarios, see~\cite{Bronnikov:2022ini} for a recent status report on
possible spatial and temporal variations of the electromagnetic
fine-structure constant $\alpha$, the gravitational constant $G$ and
the proton magnetic moment $\mu_p$. See also the recent
review~\cite{Gupta:2023jcm}.

As is well known, primordial or Big Bang nucleosynthesis (BBN) is a
fine laboratory to test our understanding of the fundamental physics
describing the generation of the light elements. In particular, it
sets bounds on the possible variation of the parameters of the
Standard Model of particle physics as well as the Standard Model of
cosmology ($\Lambda$CDM). For recent reviews, see {\it e.g.}
Refs.~\cite{Iocco:2008va,Uzan:2010pm,Adams:2019kby}. Here, we are
interested in bounds on the electromagnetic fine-structure constant
$\alpha$ derived from the element abundances in primordial
nucleosynthesis.  For earlier work on this topic, see {\textit{e.g.}
\cite{Bergstrom:1999wm,Nollett:2002da,Coc:2006sx,Dent:2007zu} and
references therein. This work is part of a larger program that tries
to map out the habitable universes in the sense that the pertinent
nuclei needed to generate life as we know it are produced in the Big
Bang and in stars in a sufficient amount, see {\it e.g.}
\cite{Meissner:2014pma,Donoghue:2016tjk} for reviews.

Here, we focus largely on the nuclear and particle physics underlying
the element generation in primordial nucleosynthesis. In particular,
we reassess the dependence of the nuclear reactions rates on the
fine-structure constant, overcoming on one side certain approximations
made in the literature and on the other side providing new and
improved parameterizations for the most important reactions in the
reaction network, using modern determinations of the ingredients
whenever possible, such as the Effective Field Theory (EFT)
description of the leading nuclear reaction $n+p\to d+\gamma$~\footnote{%
There are some {\em ab initio} calculations of other reactions in
the BBN network such
as~\cite{Marcucci:2015yla,Dohet-Eraly:2015ooa,Higa:2016igc,Premarathna:2019tup,Hebborn:2022iiz,Higa:2022mlt},
  mainly concerned with radiative capture reactions. The calculations
  in the framework of so-called ``halo-EFT'' potentially offer the
  possibility to study the $\alpha$ dependence of the cross sections
  analytically, but the implementation is numerically rather involved and we thus
  refrained from doing so in the present context. 
} 
and the
calculation of the nuclear Coulomb energies based on Nuclear Lattice
Effective Field Theory. For $\beta$-decays, we also use up-to-date
information on the neutron-proton mass difference based on dispersion
relations (Cottingham sum rule).  Most importantly, as already done in
Ref.~\cite{Meissner:2022dhg}, we utilize four different publicly
available codes for BBN
\cite{Kawano:1992ua,Arbey:2011nf,Arbey:2018zfh,Pisanti:2007hk,%
  Consiglio:2017pot,Gariazzo:2021iiu,Pitrou:2018cgg}
to address the systematic uncertainties related to the modeling of the
BBN network. In particular these codes differ in the number of nuclei
and reactions taken into account as well as in the specific
parameterization of the nuclear rates entering the coupled rate
equations for the BBN network. Moreover, in determining the
sensitivity of primordial abundances on nuclear parameters, we account
for the temperature dependence of the variation of some rates on the
value of the fine-structure constant $\alpha$.  To our knowledge, such
a comparative study where this temperature dependence was explicitly
considered has not been published before. We further note that we keep
all other constants, like {\em e.g.} the light quark masses $m_u,m_d$
fixed at their physical values.

The paper is organized as follows: In Sect.~\ref{sec:bf} we collect
the basic formulas needed for discussing the fine-structure constant
dependence in BBN. In this section we discuss the various dependences
of the reaction rates on the value of the electromagnetic
fine-structure constant $\alpha$. The actual calculation of the
reaction rates is treated in Sect.~\ref{sec:crearates}. The BBN
response matrix is introduced in Sect.~\ref{sec:response}. The
numerical results of this study are presented in
Sect.~\ref{sec:results} and discussed in Sect.~\ref{sec:summary}.  We
also present a detailed comparison to results obtained in earlier
works. In Appendix~\ref{app:params} we give the novel
parameterizations of 18 leading nuclear reactions in the BBN network.

\section{\label{sec:bf}Basic formalism}

As discussed in Ref.~\cite{Meissner:2022dhg}, the basic quantities to
be determined in BBN are the nuclear abundances $Y_{n_i}$, where $n_i$
denotes some nuclide. The evolution of the nuclear abundance $Y_{n_1}$ is
then generically given by
\begin{eqnarray}
  \dot Y_{n_1}
  &=&
      \!\!\!\!\!\!\!\sum_{\footnotesize
      \begin{array}[c]{c}
        n_2,\ldots,n_p\\
        m_1,\ldots,m_q\\
      \end{array}
  }
  \!\!\!\!\!\!\!N_{n_1}
  \Biggl(
  \Gamma_{m_1,\ldots,m_q \to n_1,\ldots n_p}
  \frac{Y_{m_1}^{N_{m_1}}\!\cdots\!Y_{m_q}^{N_{m_q}} }{
  N_{m_1}!\cdots N_{m_q}!}
  \nonumber\\ 
  &&
     \quad - \quad
     \Gamma_{n_1,\ldots,n_p \to m_1,\ldots m_q}
  \,\frac{Y_{n_1}^{N_{n_1}}\cdots Y_{n_p}^{N_{n_p}} }{
     N_{n_1}!\cdots N_{n_p}!}
     \Biggr)\,,\quad
\end{eqnarray}
where the dot denotes the time derivative in a comoving frame, and $N_{n_i}$
is the stochiometric coefficient of species $n_i$ in the reaction. Further,
 for a two-particle reaction $a + b \to c + d$\,,
$\Gamma_{ab \to cd} = n_B \gamma_{ab\to cd}$ is the reaction rate with
$n_B$ the baryon volume density.  This can readily be generalised to
reactions involving more (or less) particles,
see~\cite{Pitrou:2018cgg}. These equations are coupled via
the corresponding energy densities to the standard Friedmann equation
describing the cosmological expansion in the early universe, for
details and basic assumptions, see
also~\cite{Pitrou:2018cgg,Arbey:2018zfh,Gariazzo:2021iiu}.
In what follows, we discuss the various types of reactions in the BBN
network and their dependence on the electromagnetic fine-structure
constant.

\subsection{\label{subsec:nrr}Reaction rates}

The average reaction rate $\gamma_{ab\to cd} = N_A\,\langle
\sigma_{ab\to cd}\,v\rangle$ for a two-particle reaction $a + b \to c
+ d$ is obtained by folding the cross section $\sigma_{ab\to cd}(E)$
with the Maxwell-Boltzmann velocity distribution in thermal
equilibrium
\begin{equation}
  \label{eq:crossrate}
  \gamma_{ab\to cd}(T) = N_A\,
  \sqrt{\frac{8}{\pi \mu_{ab} (kT)^3}}\,
  \intdif{0}{\infty}{E}\,E\,\sigma_{ab\to cd}(E)\,\expo{-\frac{E}{kT}},
\end{equation}
conventionally multiplied by Avogadro's number $N_A$\,, where
$\mu_{ij}$ is the reduced mass of the nuclide pair $ij$,
$\mu_{ij}=m_im_j/(m_i+m_j)$, $E$ is the
kinetic energy in the center-of-mass system (CMS), $T$ is the
temperature and $k$ the Boltzmann constant.  Defining $y = {E}/(k T)$
this can be written in the form 
\begin{equation}
  \gamma _{ab\to cd}(T)
  = N_A
  \sqrt{\frac{8\,k T}{\pi\,\mu_{ab}}}
  \intdif{0}{\infty}{y}\,\sigma_{ab\to cd}(k Ty)\,y\,\expo{-y}~.
\end{equation}
This is suited for numerical computation \textit{e.g.} with a
Gau{\ss}-Laguerre integrator. In fact, in order to deal with cases
with singular cross sections for $E \to 0$ it is even better to split
the integral and write
\begin{eqnarray}
  \label{eq:splitint}
  &&\intdif{0}{\infty}{y}\,\sigma(kTy)\,y\,\expo{-y}
     \nonumber\\
  &=&
      2\,\intdif{0}{\sqrt{\overline{y}}}{x}\,\sigma(kTx^2)\,
      x^3\,\expo{-x^2}
      +
      \intdif{\overline{y}}{\infty}{y}\,\sigma(kTy)\,
      y\,\expo{-y}
      \nonumber
  \\
  &&
\end{eqnarray}
and evaluate the first integral with a Gau{\ss}-Legendre and the
second with a Gau{\ss}-Laguerre integrator for some suitable value of
$\overline{y}$. Note that in the first term the substitution
$x=\sqrt{y}$ was performed.

With the detailed balance relation
\begin{equation}
  \sigma_{cd\to ab}(E') 
  =
  \frac{g_a\,g_b}{g_c\,g_d}\,\frac{p^2}{{p'}^2}\,\sigma_{ab\to cd}(E)\,,
\end{equation}
where 
\begin{equation}
E = \frac{p^2}{2\,\mu_{ab}}\,, ~~~E'=\frac{{p'}^2}{2\,\mu_{cd}}~,
\end{equation}
are the CMS kinetic energies in the entrance and exit channels,
respectively, and $g_i$ is the spin multiplicity of particle $i$, 
energy conservation implies
\begin{eqnarray}
  m_a + m_b + E
  &=&
      m_c + m_d + E'
      \quad
      \textnormal{or}
      \quad
      \nonumber  \\
  E' = E + Q\,,&\textnormal{ with }&
                                   Q = m_a+m_b-m_c-m_d~,
\end{eqnarray}
in terms of the $Q$-value for the forward reaction. In thermal
equilibrium the inverse reaction rate is then related to the forward
rate as
\begin{equation}
  \label{eq:inverserate}
  \gamma_{cd \to ab}(T)
  =
  \left(\frac{\mu_{ab}}{\mu_{cd}}\right)^{\frac{3}{2}}
  \frac{g_a\,g_b}{g_c\,g_d}\, 
  \expo{-\frac{Q}{k T}}\,
  \gamma_{ab\to cd}(T)\,.
\end{equation}

This brings us to the central question of this paper, namely how the
value of the electromagnetic fine-structure constant
\begin{equation}
  \alpha = \frac{e^2}{\hbar c}
\end{equation}
influences the reaction rates? This clearly depends on the reaction
type. With the exception of the leading $n + p \to d + \gamma$ nuclear
reaction, to be discussed in some detail below, no \textit{ab initio}
expressions for most of the reaction cross sections is available and
accordingly one has to rely on model assumptions concerning the
fine-structure constant dependence of the cross sections and thus of
the reaction rates (see also the discussion in Sect.~\ref{sec:summary}
on this issue).  These shall be discussed in the following subsections
separately for {\em direct reactions} of the type
\begin{equation}
  a+b \to c+d~,
\end{equation}
{\em radiative capture reactions} of the type 
\begin{equation}
  a+b \to c + \gamma~,
\end{equation}
and {\em $\beta$-decays}, 
\begin{equation}
  a \to c + e^\mp + \stackrel{(-)}{\nu}.
\end{equation} 
We shall start with a brief discussion of the Coulomb penetration
factor for charged particles, relevant for what follows.

\subsubsection{\label{sec:penetration}{Coulomb-penetration factor}}

The Coulomb-penetration factor for an $l$-wave is given by, see
\textit{e.g.}~\cite{Humblet:1964XYZ,Jeukenne:1964XYZ}\,,
\begin{equation}
  v_\ell(\eta,\rho) =
  \frac{1}{F_\ell^2(\eta,\rho)+G_\ell^2(\eta,\rho)}\,,
\end{equation}
where $F_\ell\,,G_\ell$ are the regular and irregular Coulomb
functions, respectively, that are the linearly independent solutions
of the radial Schr\"odinger equation
\begin{equation}
  u_\ell''(\rho) + \left(1-\frac{2\,\eta}{\rho}
    -
    \frac{\ell(\ell+1)}{\rho^2}\right)\,u_\ell(\rho) = 0\,,
\end{equation}
where we defined
\begin{eqnarray}
  k
  &=&
      \sqrt{\frac{2\,\mu_{ab} c^2\,E}{\hbar^2 c^2}}\,,
      \nonumber\\
  \rho &=& k\,r\,,
  \nonumber\\
  \eta &=& \frac{Z_a\,Z_b\,\mu_{ab} c^2\,\alpha}{\hbar c\,k}\,,
\end{eqnarray}
for the Coulomb-scattering of charges $Z_a e\,,\,Z_b e$ with masses
$m_a\,,m_b$  and the reduced mass 
\begin{equation}
\mu_{ab}=\frac{m_a\,m_b}{m_a+m_b}
\end{equation}
at the energy $E$ of the relative motion, subject to the
Coulomb-potential 
\begin{equation}
V(r) =
\frac{Z_a\,Z_b\,e^2}{r}=\frac{Z_a\,Z_b\,\alpha\,\hbar c}{r}~.
\end{equation}
Approximate parameterizations of $v_{\ell}(\eta,\rho)$ have been
extensively discussed in the literature, see~ \textit{e.g.}
\cite{Humblet:1987XYZ} in particular for the dependence on the nuclear
distance $r$ where this is to be evaluated for a specific reaction.
As argued in~\cite{Humblet:1964XYZ,Jeukenne:1964XYZ}, this distance is
not well defined and the cross section should not depend on such an
unobservable parameter. Accordingly, if one takes, as
in~\cite{Humblet:1964XYZ,Jeukenne:1964XYZ}, $\lim \rho\to 0$ the
penetration factor for an $\ell$-wave then reads
\begin{eqnarray}
  \label{eq:vleta}
  v_\ell(\eta) &\approx& \varepsilon_\ell^2\,, \nonumber\\
  \varepsilon_\ell^2 &=&
  \left(1+\frac{\eta^2}{\ell^2}\right) \varepsilon_{\ell-1}^2\, , \nonumber\\
  \varepsilon_0^2 &=& \frac{2\pi\,\eta}{\expo{2\pi\,\eta}-1}\,.
\end{eqnarray}

Therefore, we shall use as Coulomb penetration factor the expression
for $s$-waves:
\begin{equation}
  \label{eq:penetration}
  P(x) = \frac{x}{\expo{x}-1}\,,
  \quad
  \lim_{x\to 0}P(x) = 1~.
\end{equation}
Note that the corrections due to a variation of $\alpha$ in
$\varepsilon_\ell^2$ for $\ell>0$ according to Eq.~(\ref{eq:vleta})
are of higher order in $\alpha$ and thus small anyway\,. Here we
defined
\begin{equation}
  \label{eq:eta}
  x = 2\pi\,\frac{Z_a\,Z_b\,\mu_{ab} c^2\,\alpha}{\hbar c\,k}
  = \sqrt{\frac{E_G(\alpha)}{E}}
\end{equation}
in terms of the so-called Gamow energy for a two-particle reaction channel $ij$
\begin{equation}
  \label{eq:GamowE}
E_G(\alpha) = 2\,\pi^2\,Z_i^2\,Z_j^2\,\mu_{ij}^{} c^2\,\alpha^2
\end{equation}
and the CMS energy $E$ or $E+Q$ for the entrance and exit channel,
respectively.

\subsection{\label{sec:direct}Direct reactions $a+b \to c + d$}

For a direct reaction of this type the $Q$-value is given by 
\begin{equation}
  Q = m_a + m_b - m_c - m_d~,
\end{equation}
where the nuclear mass of each nuclide $i$ with mass number $A_i$ and
charge number $Z_i$ reads
\begin{equation}
  m_i = Z_i\,m_p + (A_i-Z_i)\,m_n - B_i\,,
\end{equation}
with $B_i$ the nuclear binding energy. Thus, because of baryon number
and charge conservation ($A_a+A_b=A_c+A_d\,, Z_a + Z_b = Z_c +
Z_d$)\,:
\begin{equation}
  Q = B_c+B_d-B_a-B_b\,.
\end{equation}
Now the binding energy can be written as
\begin{equation}
  B_i = B_i^N - V_i^C(\alpha)\,,
\end{equation}
where $B_i^N$ denotes the strong contribution to the binding energy
and
\begin{equation}
  V_i^C(\alpha) \propto \alpha\,Z_i\,(Z_i-1)
\end{equation}
is the expectation value of the Coulomb contribution proportional to
the value of the electromagnetic fine-structure constant. Considering
its variation in the form $\alpha = \alpha_0\,(1+\delta_\alpha)$\,,
where
\begin{equation}
  \alpha_0
  =
  7.297 3525693(11)\,10^{-3}
  =
  1/137.035 999 084(2)
\end{equation}
is the current experimental value from Ref.~\cite{Workman:2022ynf},
the $Q$-value varies as
\begin{equation}
  \label{eq:Qabcd}
  Q(\alpha) = Q(\alpha_0) + (V^C_a+ V^C_b - V^C_c - V^C_d)\,\delta_\alpha\,.
\end{equation}
One therefore needs an estimate of the Coulomb contribution to the
nuclear masses, this we shall discuss in
Sect.~\ref{sec:Coulombenergies}.

We shall assume that the cross section for a direct reaction $a+b \to
c + d$ depends on $\alpha$ as
\begin{eqnarray}
  \label{eq:sigmaabcd}
  \lefteqn{
  \sigma_{ab\to cd}\left(E;Q(\alpha),E_G^i(\alpha),E_G^i(\alpha)\right)
  }
  \nonumber\\
  &=&
      \sqrt{E+Q(\alpha)}\,P_i(x_i(E,\alpha))\,P_f(x_f(E,\alpha))\,f(E)~,
\end{eqnarray}
where $f$ is some function independent of $\alpha$ and $P_i(x_i)$,
$P_f(x_f)$ are the penetration factors given by
Eq.~(\ref{eq:penetration}) reflecting the Coul{\-}omb repulsion in the
entrance and in the exit channel, respectively.  The first factor in
Eq.~(\ref{eq:sigmaabcd}) accounts for the exit channel momentum
dependence of the cross section of the direct reaction $a + b \to c +
d$.  Here,
\begin{eqnarray}
  \label{eq:pargi}
  x_i(E,\alpha) &=& \sqrt{\frac{E_G^i(\alpha)}{E}}\,,
  \label{eq:pargf}
  \\
  x_f(E,\alpha) &=& \sqrt{\frac{E_G^f(\alpha)}{E+Q(\alpha)}}\,,
\end{eqnarray}
are the arguments of the penetration factors, with
\begin{eqnarray}
  \label{eq:GamowEi}
  E_G^i(\alpha)
  &=&
      2\,\pi^2\,Z_a^2\,Z_b^2\,\mu_{ab}^{} c^2\,\alpha^2\,,
  \\
  \label{eq:GamowEf}
  E_G^f(\alpha)
  &=&
      2\,\pi^2\,Z_c^2\,Z_d^2\,\mu_{cd}^{} c^2\,\alpha^2\,,
\end{eqnarray}
the Gamow-energies in the entrance and the exit channel, respectively,
and $\mu_{ij}=m_i\,m_j/(m_i+m_j)$ the corresponding reduced masses.
Although in order to calculate the linear response of the abundances
one could proceed by calculating first order partial derivatives
\begin{equation}
  \diff{\sigma}{\alpha}
  =
  \diff{\sigma}{Q}\,\diff{Q}{\alpha}+
  \sum_{k=i,f}\diff{\sigma}{P_k}\,\diff{P_k}{x_k}\,\diff{x_k}{\alpha}\,,
\end{equation}
\textit{etc.}\,, we prefer not to presume linearity and rather
calculate a variation of the cross section with a variation $\alpha =
\alpha_0\,(1+\delta_\alpha)$ through the expression
\begin{eqnarray}
  \label{eq:sigmavar}
  \lefteqn{
  \sigma_{ab\to cd}\left(E;Q(\alpha),E_G^f(\alpha),E_G^i(\alpha)\right)
  }
  \nonumber\\
&&
   =
   \sigma_{ab\to cd}\left(E;Q(\alpha_0),E_G^i(\alpha_0),E_G^f(\alpha_0)\right)\,
   \\
&&\times
   \frac{P(x_i(E,\alpha))}{P(x_i(E,\alpha_0))}\,
   \sqrt{\frac{E+Q(\alpha)}{E+Q(\alpha_0)}}\,
   \frac{P(x_f(E,\alpha))}{P(x_f(E,\alpha_0))}\,,
   \nonumber
\end{eqnarray}
where specifically the first factor reads
\begin{eqnarray}
  \label{eq:inifac}
  \frac{P(x_i(E,\alpha))}{P(x_i(E,\alpha_0))}
  &=&
      \frac{\sqrt{\frac{E_G^i(\alpha)}{E}}}{\expo{\sqrt{\frac{E_G^i(\alpha)}{E}}}-1}\,
      \frac{\expo{\sqrt{\frac{E_G^i(\alpha_o)}{E}}}-1}{\sqrt{\frac{E_G^i(\alpha_0)}{E}}}
      \nonumber\\
  &&\hspace*{-5em}=
     \left\lbrace
     \begin{array}[c]{ll}
       1, &
            \textnormal{for $n$-induced reactions}
       \\
       (1+\delta_\alpha)\,
       \frac{\expo{\sqrt{\frac{E_G^i(\alpha_o)}{E}}}-1}{\expo{\sqrt{\frac{E_G^i(\alpha)}{E}}}-1}\,,
          & \textnormal{else,}\\
     \end{array}
  \right.
\end{eqnarray}
and the remaining factors are given by
\begin{eqnarray}
  \label{eq:finfac}
  \lefteqn{
  \sqrt{\frac{E+Q(\alpha)}{E+Q(\alpha_0)}}\,
  \frac{P(x_f(E,\alpha))}{P(x_f(E,\alpha_0))}
  }
  \nonumber\\
&&\qquad=
   \left\lbrace
   \begin{array}[c]{ll}
     \sqrt{\frac{E+Q(\alpha)}{E+Q(\alpha_0)}}\,,
     &
       \textnormal{if $c=n$ and/or $d=n$}
     \\
     \sqrt{\frac{E_G^f(\alpha)}{E_G^f(\alpha_0)}}\,
     \frac{\expo{\sqrt{\frac{E_G^f(\alpha_0)}{E+Q(\alpha_0)}}}-1}{
     \expo{\sqrt{\frac{E_G^f(\alpha)}{E+Q(\alpha)}}}-1}\,,
     &
       \textnormal{else.}\\
   \end{array}
  \right.
\end{eqnarray}
We note that these factors are energy-dependent and therefore the
change in the rate
\begin{eqnarray}
  \label{eq:ratevar}
  \lefteqn{
  \gamma(T;Q(\alpha),E_G^i(\alpha),E_G^f(\alpha))
  }
  \nonumber\\
  &=&
  \intdif{0}{\infty}{E}\,\,E\,
  \sigma(E;Q(\alpha),E_G^i(\alpha),E_G^f(\alpha))\,\,
      \expo{-\frac{E}{kT}}\,,
\end{eqnarray}
\textit{i.e.} the factor
\begin{equation}
  \frac{
    \gamma(T;Q(\alpha),E_G^i(\alpha),E_G^f(\alpha))
  }{
    \gamma(T;Q(\alpha_0),E_G^i(\alpha_0),E_G^f(\alpha_0))
  }
\end{equation}
depends on the temperature $T$ and as it stands requires a numerical
evaluation of Eq.~(\ref{eq:ratevar})\,.

\subsection{\label{sec:radcap}Radiative capture reactions $a+b \to c + \gamma$}

Similar considerations hold for radiative capture reactions. The cross
section of a reaction $a+b \to c + \gamma$ is assumed to depend on
$\alpha$ as
\begin{eqnarray}
  \label{eq:sigmaradcap}
  \lefteqn{
  \sigma_{ab\to c\gamma}(E;Q(\alpha),E_G^i(\alpha))
  }
  \nonumber\\
  &=&
      \alpha\,(E+Q(\alpha))^3\,P_i(x_i(E,\alpha))\,f(E)\,
\end{eqnarray}
with $f$ some $\alpha$-independent function and $P_i(x_i)$ the
penetration factor, see
Eqs.~(\ref{eq:penetration},\ref{eq:pargi},\ref{eq:GamowEi}), for the
entrance channel.  The first factor accounts for the fact that in the
amplitude for a radiative capture reaction the photon coupling is
proportional to $e$, leading to a factor proportional to $\alpha =
e^2/(\hbar c)$ in the cross section.  The second factor reflects the
final momentum dependence assuming dipole dominance of the radiation%
\footnote{%
  Note, however, that this is not always the case, exceptions with
  appreciable $E2$ (electric quadrupole) contributions are
  \textit{e.g.} the reactions:
  ${^2}\textnormal{H}+{^2}\textnormal{H} \to {^4}\textnormal{He} + \gamma$\,,
  ${^2}\textnormal{H}+{^4}\textnormal{He} \to {^6}\textnormal{Li} + \gamma$
  and
  ${^4}\textnormal{He}+{^{12}}\textnormal{O} \to {^{16}}\textnormal{O} + \gamma$\,.
  We nevertheless always assume dipole dominance.
}%
.
We thus calculate a variation of the cross section for radiative
capture with a variation $\alpha = \alpha_0\,(1+\delta_\alpha)$ via
\begin{eqnarray}
  \label{eq:sigmavarg}
  \lefteqn{
  \sigma_{ab\to c\gamma}\left(E;Q(\alpha),E_G^i(\alpha)\right)
  }
  \nonumber\\
&&
   =
   \sigma_{ab\to c\gamma}\left(E;Q(\alpha_0),E_G^i(\alpha_0)\right)\,
   \nonumber\\
&&\times
   \frac{P(x_i(E,\alpha))}{P(x_i(E,\alpha_0))}\,
   (1+\delta_\alpha)\,
   \left(\frac{E+Q(\alpha)}{E+Q(\alpha_0)}\right)^3
\end{eqnarray}
where the first factor is the same as in Eq.~(\ref{eq:inifac})\,.
Again note that both factors are energy-dependent and therefore a
change in the rate, see Eq.~(\ref{eq:ratevar}), is
temperature-dependent.

\subsection{\label{sec:approx}Approximate treatment of $\alpha$-dependent factors}

As mentioned twice, the variation of the cross sections with a
variation of $\alpha$ induces energy-dependent factors, that in turn
lead to temperature-dependent variations in the corresponding reaction
rates, that can be fully accounted for only via a numerical
integration of Eq.~(\ref{eq:ratevar}). In fact this is what was done
in the present work for the most important 18 nuclear reactions in the
BBN network, listed in Sect.~\ref{sec:results}.  For the remaining
reactions we relied on the following approximations, that turned out
to be effective.
\begin{figure*}[!htb]
  \includegraphics[width=\textwidth, trim=40 230 73 80, clip]{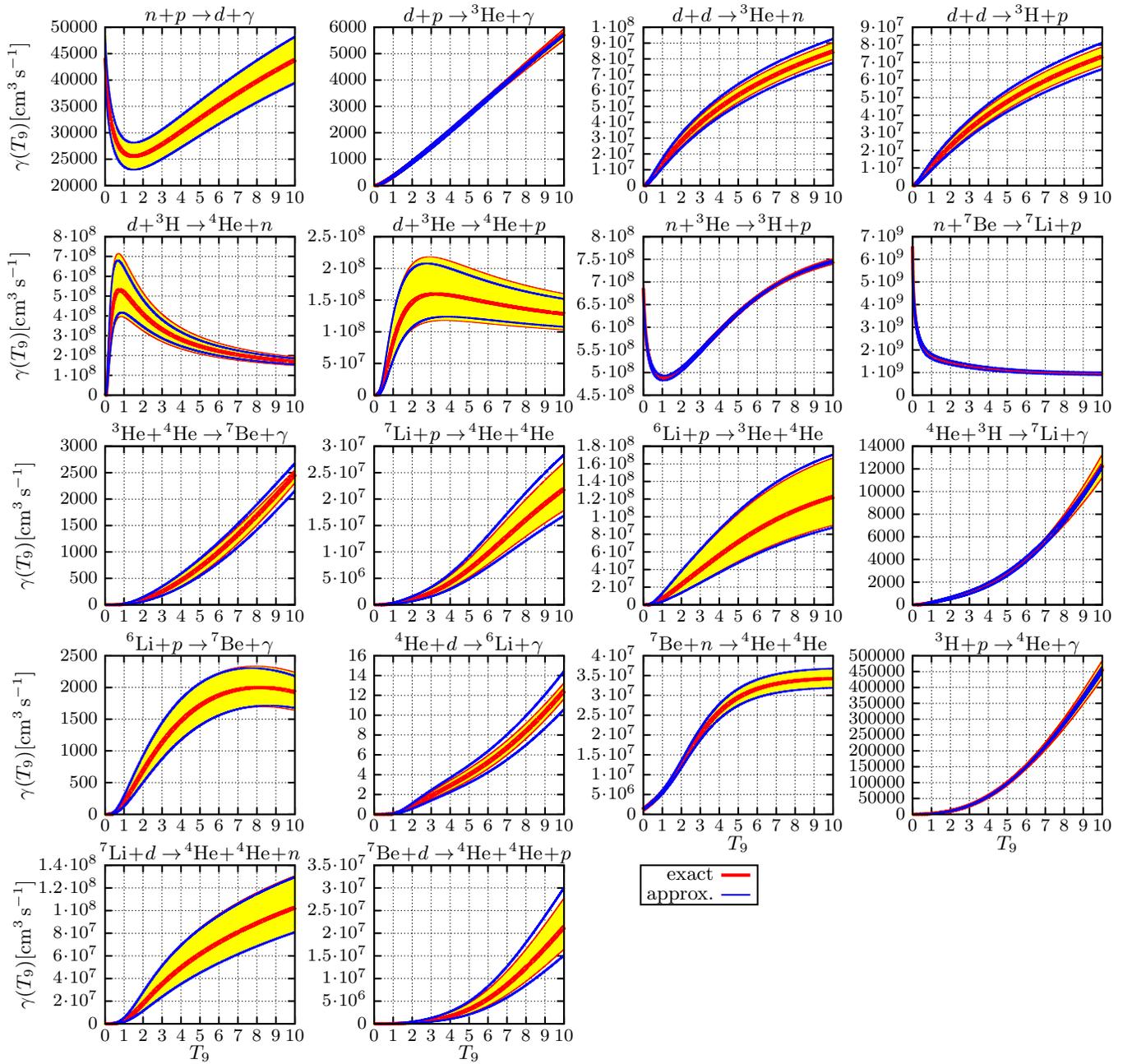}
  \caption{\label{fig:ratealpha}
    Temperature-dependence ($T_9 = T / 10^9~\textnormal{K}$) of the
    variation of the rates $\gamma$ of 18 leading nuclear reactions with a
    variation of the fine structure constant $\alpha =
    \alpha_0\,(1+\delta_\alpha)$. Shown are the exact results in the
    interval $\delta_\alpha = [-0.1,0.1]$ (yellow area, color online)
    bounded by the curves (in red) at $\delta_\alpha = -0.1$ and $0.1$ as
    well as the approximate expression discussed in Sect.~\ref{sec:approx}
    for $\delta_\alpha = -0.1$ and $0.1$ (blue curves)\,.
  }
\end{figure*}
For neutron induced reactions
\begin{equation}
  \label{eq:ninduced}
  \sigma(E) \propto \frac{R(E)}{\sqrt{E}}
\end{equation}
where for a non-resonant reaction $R(E)$ is a weakly dependent
function of the CMS kinetic energy $E$, see \textit{e.g.}
\cite{Serpico:2004gx}\,.  If we make the extreme approximation that
$R(E) \approx \textit{const.}$ the maximum of the remaining energy
dependent factors in the integrand of Eq.~(\ref{eq:ratevar}) is
reached at the energy
\begin{equation}
  \label{eq:En}
  E = \overline{E}_n = \frac{1}{2}\,kT\,.
\end{equation}
Likewise, assuming that for the astrophysical $S$-factor for charged
particle induced reactions
\begin{equation}
  \label{eq:chargereac}
  S(E) = E\,\sigma(E)\,\expo{\sqrt{\frac{E_G^i}{E}}} \approx \textit{const.}
\end{equation}
holds, one finds that the maximum of the remaining energy-dependent
factors in the rate is reached at
\begin{equation}
  \label{eq:Ec}
  E
  =
  \overline{E}_c
  =
  \left(\frac{kT}{2}\right)^{\frac{2}{3}}(E_G^i)^{\frac{1}{3}}\,.
\end{equation}
Substituting $E \mapsto \overline{E}_n, \overline{E}_c$ in the
expressions in Eqs.~(\ref{eq:sigmavar}, \ref{eq:sigmavarg}) then leads
to temperature-dependent factors, that can be taken in front of the
integral in Eq.~(\ref{eq:ratevar}) and thus merely multiply the
corresponding rates.  The quality of this approximation may be
inferred from Fig.~\ref{fig:ratealpha}, where we compare the results
of the numerical calculation of the rates according to
Eq.~(\ref{eq:ratevar}) (yellow areas for a variation $\delta_\alpha
\in [-0.1,0.1]$) with the approximation discussed in this subsection,
represented by blue lines for $\delta_\alpha = -0.1$ and $0.1$\,.

Note that in the present treatment we preferred to account for the
Coulomb suppression in an entrance or exit channel with charged
particles by the penetration factor of Eq.~(\ref{eq:penetration}) and
do not rely on a simple Gamow-factor $\propto \expo{-x} =
\expo{-\sqrt{E_G(\alpha)/E}}$, with $E$ being the CMS energy of the
relevant channel with charged particles. We found that doing so would
lead to overestimating the $\alpha$-dependence in the rates by a
factor~$\approx 1.5$, while the temperature dependence would still
roughly follow the same trends as in Fig.~\ref{fig:ratealpha}.

\subsection{\label{sec:beta}Coulomb-effects in $\beta$-decays}

Next, we consider the various $\beta$-decays in the BBN network.
The rate for $\beta$-decays $a \to c + e^{\pm} + \stackrel{(-)}{\nu}$
can be written as, see \textit{e.g.}~\cite{es1977}, 
\begin{equation}
  \label{eq:betarate}
  \lambda
  =
  \frac{G^2}{2\pi^3}\,\frac{m_p\,c^2}{\hbar}
  \left| \mathcal{M}_{ac}\right|^2\,f(Z,q)\,,
\end{equation}
where $G$ is Fermi's weak coupling constant, $\mathcal{M}_{ac}$ the
nuclear matrix element and
\begin{equation}
  \label{eq:fZq}
  f(\pm Z,q)
  =
  \intdif{1}{q}{x}\,\,F(\pm Z,x)\,\sqrt{x^2-1}\,x\,(q-x)^2\,,
\end{equation}
see Eq.~(2.158) of Ref.~\cite{morita1973}\,, where we defined $q = Q/m_e =
(m_a-m_c)/m_e$. Further, $F(\pm Z,E)$ is the so-called Fermi-funct{\-}ion
\begin{eqnarray}
  &&
     F(\pm Z, E)
     =
      F_0(\pm Z, E)\,L_0(\pm Z, E)\,,
 \nonumber \\
  &&
     L_0(\pm Z, E)
     =
      \frac{1+\gamma}{2}
      \mp
      \frac{5}{3}\,\alpha\,Z\,R\,E
      \mp
     \frac{\alpha\,Z\,R}{3\,E}
  \nonumber\\
  &&\qquad\qquad\qquad
      -
      \frac{1}{3}\,(E^2-1)\,R^2
      +
      \cdots\,,
  \nonumber\\
  &&
     F_0(\pm Z, E)
  \nonumber\\
  &&
  \quad=
      4\,(2\sqrt{E^2-1}\,R)^{2\gamma-2}\,
      \expo{\pm\pi\nu}
      \frac{{\left|\Gamma(\gamma\pm\ii\,\nu)\right|}^2}{(\Gamma(2\gamma+1))^2}
\end{eqnarray}
with the definitions
\begin{eqnarray}
  \gamma&=&\sqrt{1-(Z\,\alpha)^2}\,,
  \nonumber\\
  \nu &=& \frac{Z\,\alpha\,E}{\sqrt{E^2-1}}\,,
\end{eqnarray}
where $Z$ is the atomic number of the daughter nucleus $c$ and $R$ its
radius, see Eqs.~(2.121)-(2.125),(2.131) of Ref.~\cite{morita1973}\,.
The upper/lower sign holds for $\beta^-/\beta^+$ decays, respectively.
For $Z\,\alpha \ll 1$ we can approximate
\begin{eqnarray}
  &&
     \gamma \approx 1\,,
     \quad
     L_0(\pm Z,E) \approx 1\,,
 \nonumber \\
  &&
     F_0(\pm Z,E)
     \approx
     4\cdot 1 \cdot\,\expo{\pm\pi\nu}\,
     \frac{{\left|\Gamma(1\pm\ii\,\nu)\right|}^2}{4}
\end{eqnarray}
or, with
${\left|\Gamma(1\pm\ii\,\nu)\right|}^2 = \pm\pi\nu / \sinh{(\pm\pi\nu)}$
\begin{equation}
  \label{eq:F0}
  F_0(\pm Z,E) \approx
  \frac{\pm2\pi\,\nu\,
    \expo{\pm\pi\nu}}{\expo{\pm\pi\nu}-\expo{\mp\pi\nu}}
  =
  \frac{\pm2\pi\,\nu}{1-\expo{\mp\,2\pi\,\nu}}\,,
\end{equation}
such that $\lim_{Z\to 0} F_0(\pm Z,E) = 1$.
Accordingly, setting $a=2\pi\,Z\,\alpha$ then
\begin{eqnarray}
  \label{eq:fZqa}
  f(\pm Z,q)
  &=&
      \intdif{1}{q}{x}\,\,
      \frac{\pm \frac{a\,x}{\sqrt{x^2-1}}}{1-\expo{\mp\frac{a\,x}{\sqrt{x^2-1}}}}\,
      \sqrt{x^2-1}\,x\,(q-x)^2\,
      \nonumber \\
  &=&
  \intdif{1}{q}{x}\,\,
  \frac{\pm a\,x^2}{1-\expo{\mp\frac{a\,x}{\sqrt{x^2-1}}}}\,
  (q-x)^2\,.
\end{eqnarray}
Defining also $p = \sqrt{q^2-1}$ (\textit{i.e.} the maximal momentum
in $\beta$-decay divided by the electron mass) and with the
substitution $y=\sqrt{x^2-1}/p$ we can rewrite the expression for
$f(\pm Z,q)$ as
\begin{eqnarray}
  \label{eq:fZp}
  f(\pm Z,p)
  &=&
      f(\pm a,p)
     \nonumber \\
  &=&
      \pm a\,p^2\,\intdif{0}{1}{y}\,\,
      y\,\frac{\sqrt{1+p^2\,y^2}}{1-\expo{\mp\frac{a\,\sqrt{1+p^2\,y^2}}{p\,y}}}\,
      \nonumber\\
  &&\qquad\quad \times
     \left(\sqrt{1+p^2}-\sqrt{1+p^2\,y^2}\right)^2,
\end{eqnarray}
which is slightly better suited for a numerical implementation,
\textit{e.g} with a Gau{\ss}-Legendre-integrator. Note that both $a$
and $p$ depend on $\alpha$.

\subsubsection{\label{ssec:Dnp}Electromagnetic contribution to the proton-neutron mass difference}

The neutron-proton mass difference plays an important role in BBN, see
{\it e.g.} \cite{Hogan:1999wh}.  According to
Refs.~\cite{Gasser:2020hzn,Gasser:2020mzy} the proton-neutron mass
difference is given by
\begin{eqnarray}
  m_p - m_n &=& \Delta m =\,
                \Delta m_{\textnormal{\tiny QCD}}+\Delta m_{\textnormal{\tiny QED}}\,,
                \nonumber\\
  \Delta m_{\textnormal{\tiny QCD}}
            &=&
                -1.87 \mp 0.16\,\textnormal{MeV}\,,  
                \nonumber\\
 \Delta m_{\textnormal{\tiny QED}}
            &=&
                0.58 \pm 0.16\,\textnormal{MeV}\,,
\end{eqnarray}
where the nominal electromagnetic contribution is somewhat smaller
than the value $\Delta m_{\textnormal{\tiny QED}} = 0.7 \pm
0.3\,\textnormal{MeV}$ given earlier in Ref.~\cite{Gasser:1974wd}.  We
also note that the splitting in strong and electromagnetic
contributions is convention-dependent, for a pedagogical discussion
see~\cite{Meissner:2022odx}. For a comparison of these results
with lattice QCD and other phenomenological determinations of the
electromagnetic contribution to the neutron-proton mass difference, we
refer to~\cite{Gasser:2020hzn}.

The neutron-proton mass difference is a crucial parameter both in
the various $n \leftrightarrow p$ weak interactions in the early phase
of BBN and in all $\beta$-decays. We shall start with a discussion of
the latter.

\subsubsection{\label{sec:impbeta}Implications for $\beta$-decays}
Writing
\begin{equation}
  \label{eq:deltanp}
  m_n - m_p
  =
      -(\Delta m_{\textnormal{\tiny QCD}} + \Delta m_{\textnormal{\tiny QED}})\,,
\end{equation}
the $Q$-value for the $\beta$-decay $a \to c + e^\mp +
\stackrel{(-)}{\nu_e}$ depends on a variation of $\alpha = 
\alpha_0(1+\delta_\alpha)$ as
\begin{eqnarray}
  \label{eq:Qnp}
  Q(\alpha) 
  &=&
      Q(\alpha_0(1+\delta_\alpha))
  \nonumber\\
  &=&
      \pm(-\Delta m_{\textnormal{\tiny QCD}}
      -\Delta m_{\textnormal{\tiny QED}}\,(1+\delta_\alpha))
  \nonumber\\
  &&\qquad
     -(B^N_a-V^C_a\,(1+\delta_\alpha))
  \nonumber\\
    &&\qquad
     +(B^N_c-V^C_c\,(1+\delta_\alpha))
  \nonumber\\
  &=&
      Q(\alpha_0) + (V^C_a-V^C_c\mp\Delta m_{\textnormal{\tiny QED}})\,\delta_\alpha\,,
\end{eqnarray}
where, as in Sect.~\ref{sec:direct}, $B^N_i$ is the nuclear (strong)
contribution to the binding energy of nuclide $i$ and $V^C_i$ the
expectation value of the Coulomb-interaction to the binding energy of
nuclide $i$\,.

One thus finds for the variation of the $\beta$-decay rate with a
variation of $\alpha$,
\begin{equation}
  \label{eq:lambdabeta}
  \lambda(\alpha_0\,(1+\delta_\alpha))
  =
  \lambda(\alpha_0)\,\frac{f(\tilde a(\delta_\alpha),\tilde p(\delta_\alpha))}{f(a,p)}\,,
\end{equation}
where
\begin{eqnarray}
  \label{eq:lambdabetadef}
  \tilde a(\delta_\alpha)
  &=&
      a\,(1+\delta_\alpha)\,,
  \nonumber \\
      \tilde p(\delta_\alpha)
      &=& \sqrt{\tilde q^2(\delta_\alpha)-1}\,,
  \nonumber\\
  \tilde q(\delta_\alpha)
  &=&
      Q(\alpha_0\,(1+\delta_\alpha))/m_e~,
\end{eqnarray}
and the factor determining the variation of the $\beta$-decay rate
with a variation of $\alpha$\,, see Eq.~(\ref{eq:lambdabeta}), is
determined by evaluating $f(\tilde a,\tilde p)$ and $f(a,p)$ via
Eq.~(\ref{eq:fZp})\,.

We note that
$Q(\alpha) = Q(\alpha_0(1+\delta_\alpha)) \ge m_e\,, Q(\alpha_0) \ge m_e$
implies an upper limit for $\delta_\alpha$\,:
\begin{eqnarray}
  \hspace*{-2em}(V^C_a-V^C_c\mp\Delta m_{\textnormal{\tiny QED}})\,\delta_\alpha
  &\ge& m_e-Q(\alpha_0)
  \nonumber\\
  \Leftrightarrow
  \delta_\alpha &\le&
  \frac{Q(\alpha_0)-m_e}{V^C_c-V^C_a\pm\Delta m_{\textnormal{\tiny QED}}}\,.
\end{eqnarray}
As for other cases where during a variation of $\alpha$ the $Q$-value
of a reaction becomes negative, we have put the corresponding rate to
zero.

We also note that for the neutron decay $n \to p + e^- +
\overline{\nu}$ the variation of the rate with a variation of $\alpha$
merely implies a variation of the neutron lifetime $\tau_n \propto
1/\lambda_{n\to p}$.

\subsubsection{\label{sec:weaknp}Implications for the weak $n \leftrightarrow p$ reactions}

As detailed in Ref.~\cite{Pitrou:2018cgg} the six reactions 
\begin{eqnarray}
    \label{eq:npreac}
    n + \nu & \leftrightarrow & p + e^-\,,
                                \nonumber
    \\
    n  & \leftrightarrow & p + e^-\!+ \overline{\nu}\,,
                           \nonumber
    \\
    n + e^+ & \leftrightarrow & p + \overline{\nu}\,,
\end{eqnarray}
determine the evolution of the neutron abundance in the early phase of
BBN and hence are crucial for all other primordial nuclear
abundances. Assuming local thermodynamical equilibrium in terms of a
temperature $T$ and a distinct neutrino temperature $T_\nu$ in the
so-called infinite nucleon mass approximation the $n \to p$ (angular
averaged) reaction rate can be written, see
\textit{e.g.}~\cite{Pitrou:2018cgg} for details, as
\begin{eqnarray}
  \label{eq:Gnp}
  \lefteqn{
  \Gamma_{n \to p}(\Delta m; T)
  =
  \Gamma_{
  {n+\nu \to p\!+ e^-} \atop {n \to p + e^- + \overline{\nu}}}
  + \Gamma_{n + e^+ \to p + \overline{\nu}}
  }
  \nonumber\\
  &=& K\,\intdif{m_e}{\infty}{E}\,\,E\,\sqrt{E^2-m_e^2}
      \nonumber\\
  &&
     \times
     \Biggl[
     (E+\Delta m)^2\,g\left(\frac{E+\Delta m}{k_B T_\nu}\right)\,g\left(-\frac{E}{k_B T}\right)
     \nonumber\\
  && 
     \quad
     +
     (-E+\Delta m)^2\,g\left(\frac{-E+\Delta m}{k_B T_\nu}\right)\,g\left(\frac{E}{k_B T}\right)
     \Biggr]
\end{eqnarray}
with
\begin{equation}
  \label{eq:FDfunc}
  g(x) = \frac{1}{\expo{x}+1}
\end{equation}
the Fermi-Dirac distribution function.  The ratio $T_\nu/T$ follows
from the cosmological evolution, see the black curve in
Fig.~\ref{fig:nprate}.
\begin{figure}[htb]
  \centering
  \includegraphics[width=0.95\columnwidth, trim = 125 350 100 50, clip]{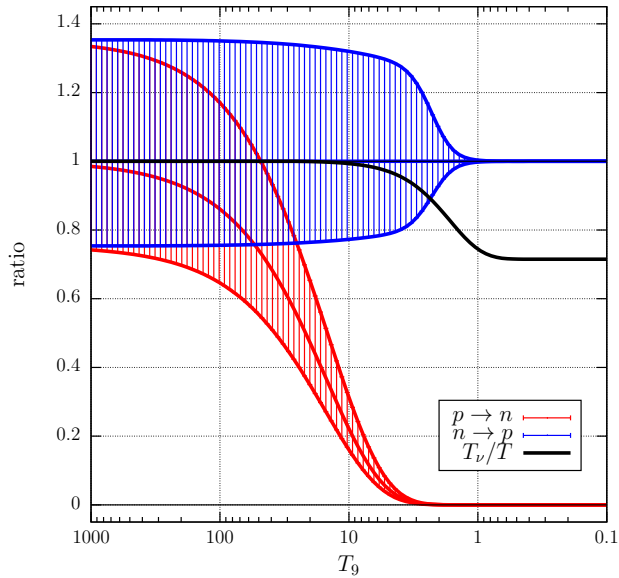}
  \caption{
    Variation of the rate ratios
    $R_{n \to p}$ (Eq.~(\ref{eq:RRnp}), blue hatched area)
    and
    $R_{p \to n}$ (Eq.~(\ref{eq:RRpn}), red hatched area)
    with decreasing temperature
    ($T_9 = T/[10^9\,\textnormal{K}]$) for $\delta_\alpha$ in the range
    $\delta_\alpha = -0.1$ (lower curves) up to $\delta_\alpha = 0.1$
    (upper curves). Also shown is the ratio $T_\nu/T$ (black curve).
  }
  \label{fig:nprate}
\end{figure}
The constant $K$ is fixed by requiring that $\Gamma_{n \to p}(\Delta
m; 0) = 1/\tau_{n}$, with $\tau_n$ the neutron lifetime. The $p \to n$
rate is simply given by substituting $\Delta m \mapsto -\Delta m$ in
Eq.~(\ref{eq:Gnp}) above. In this case $\Gamma_{p \to n}(\Delta m; 0)
= 0$\,.  As discussed in~\cite{Pitrou:2018cgg}
and~\cite{Serpico:2004gx} there are a number of corrections to the $n
\to p$ and $p \to n$ rates as given above, \textit{viz.} the Coulomb
correction (as discussed above in section~\ref{sec:impbeta}),
electromagnetic radiative corrections, finite nucleon mass
corrections, plasma corrections and non-instantaneous neutrino
decoupling effects. Some of these involve the fine-structure constant
$\alpha$, but since these effects are small corrections anyway, the
most relevant effect when varying $\alpha$ is through the change
$\Delta m(\alpha) = \Delta m(\alpha_0) -\Delta m_{\textnormal{\tiny
QED}}\,\delta_\alpha$\,. This effect is illustrated in
Fig.~\ref{fig:nprate}, where the double rate ratios
\begin{equation}
  \label{eq:RRnp}
  R_{n \to p} =
  \frac{\Gamma_{n \to p}(\Delta m(\alpha); T)\,\Gamma_{n \to p}(\Delta m(\alpha_0); 0)}{%
    \Gamma_{n \to p}(\Delta m(\alpha); 0)\,\Gamma_{n \to p}(\Delta m(\alpha_0); T)}~,
\end{equation}
and
\begin{equation}
  \label{eq:RRpn}
  R_{p \to n} = \frac{\Gamma_{p \to n}(\Delta m(\alpha); T)\,\Gamma_{n \to p}(\Delta m(\alpha_0); 0)}{%
    \Gamma_{n \to p}(\Delta m(\alpha); 0)\,\Gamma_{n \to p}(\Delta m(\alpha_0); T)}\,,
\end{equation}
obtained by a numerical integration according to Eq.~(\ref{eq:Gnp})
(with a method similar to that of Eq.~(\ref{eq:splitint})) are plotted
as a function of $T_9 = T/[10^9\,\textnormal{K}]$).  This double ratio
was chosen such that for $T \to 0$ the $n \to p$ curves tend to unity
and the $p \to n$ curves to zero; the $\alpha$-dependence of the $n
\to p$ rate in this low-temperature limit is then given by the
expressions in the preceding section~\ref{sec:impbeta}.  As is evident
from this figure the variation of the rates with varying $\alpha$ is
non-linear and strongly temperature dependent.

\subsection{\label{sec:npdg}The $n+p \to d + \gamma$ reaction}

Fortunately, for the $n + p \to d + \gamma$ reaction an accurate
treatment within the framework of pionless
EFT~\cite{Chen:1999bg,Rupak:1999rk} is available. Accordingly, for
this leading nuclear reaction in the BBN network it is possible to
study dependences of the cross section and hence of the reaction rate
on various nuclear parameters, such as the binding energy of the
deuteron, $np$~scattering lengths, effective ranges \textit{etc.} as
was done in~\cite{Meissner:2022dhg}. Here we shall focus on the
$\alpha$-dependence.

The cross section was given in~\cite{Rupak:1999rk} as
\begin{eqnarray}
  \label{eq:npdgx}
  \lefteqn{
  \sigma_{np\to d\gamma}(p)
  =
  4\pi\,\alpha
  }
  \nonumber\\
  &&
     \times
      \left(
      1-\frac{2\,p^4  + 4\,p^2\,\gamma^2 + 3\,\gamma^4}{4\,m_N^2\,(p^2+\gamma^2)}
      \right)\,
     \frac{(\gamma^2 + p^2)^3}{\gamma^3\,m_N^4\,p}
  \nonumber\\
  &&
      \times\Bigl[
      \left\langle \widetilde\chi_{E1_V}\right\rangle^2
      +
     \left\langle \widetilde\chi_{M1_V}\right\rangle^2
     +
     \left\langle \widetilde\chi_{M1_S}\right\rangle^2
     +
     \left\langle \widetilde\chi_{E2_S}\right\rangle^2
     \Bigr]\,,
\end{eqnarray}
where $p$ is the relative momentum, $m_N = (m_p+m_n)/2$ denotes the
nucleon mass, $\gamma = \sqrt{B_d\,m_N}$ is the so-called binding
momentum, with $B_d = 2.225$~MeV the binding energy of the deuteron, and
$\left\langle \widetilde\chi_{E1_V}\right\rangle^2$\,,
$\left\langle \widetilde\chi_{M1_V}\right\rangle^2$\,,
$\left\langle \widetilde\chi_{M1_S}\right\rangle^2$\,,
$\left\langle \widetilde\chi_{E2_S}\right\rangle^2$\,,
are the dimensionless amplitudes for isovector electric dipole,
isovector magnetic dipole, isoscalar magnetic dipole and iso\-scalar
electric quadrupole contributions, respectively. For the energies
relevant in BBN only the isovector contributions are significant and
these were calculated at N4LO and N2LO for the electric and magnetic
parts, respectively. The overall theoretical uncertainty is claimed to
be better than $1\%$ for CMS energies $E \le
1~\textnormal{MeV}$\,. The expression of Eq.~(\ref{eq:npdgx}) with all
terms included was used to calculate the cross section for this
reaction throughout the present investigation.

Concerning the variation of this cross section when varying
$\alpha=\alpha_0\,(1+\delta_\alpha)$ it is evident that the dominant
effect is simply
\begin{equation}
  \label{eq:npdgvar}
  \sigma_{np\to d\gamma}(\alpha;p)
  =
  (1+\delta_\alpha)\,\sigma_{np\to d\gamma}(\alpha_0;p)\,.
\end{equation}
Note that there is no Coulomb-contribution to the binding energy of
the deuteron, while the expectation value $\langle
v^{\textit{EM}}\rangle$ of the electromagnetic interaction, mainly due
to the magnetic dipole-dipole interaction moment term,
see~\cite{Wiringa:1994wb} for a treatment based on the Argonne
$v_{18}$ nucleon-nucleon potential, is very small, $\langle
v^{\textit{EM}}\rangle = 0.018~\textnormal{MeV}$.  Hence the effects
of a change of the $Q$-value of the reaction with varying $\alpha$, as
discussed in the previous subsections, are considered to be negligible
in the present context. Moreover, in the expression of
Eq.(~\ref{eq:npdgx}), as well as in the expressions for the amplitudes
$\widetilde\chi$ of Ref.~\cite{Rupak:1999rk} the nucleon mass $m_N=
(m_p+m_n)/2$ occurs at various instances. Although a moderately
accurate value for the electromagnetic contribution to the
neutron-proton mass difference is available (and was, in fact, used in
our discussion of the $\beta$-decays in Sect.~\ref{ssec:Dnp}), only
rough estimates are available for the electromagnetic contribution to
the neutron and proton mass separately. In Eq.~(12.3) of
Ref.~\cite{Gasser:1982ap} the estimates $m_p^{\textnormal{Born}}
\approx 0.63$~MeV, $m_n^{\textnormal{Born}} \approx -0.13$~MeV (with
an estimated accuracy of $\approx 0.3~\textnormal{MeV}$) for the total
electromagnetic self-energy can be found, which, via
$m_N^{\textnormal{Born}} \approx 0.25~\textnormal{MeV}$\,, would imply
that $m_N$ varies with $\alpha$ (putting $m_N \approx
1~\textnormal{GeV}$ for this estimate) as
\begin{equation}
  m_N(\alpha) \approx m_N(\alpha_0)(1+0.00025\,\delta_\alpha)\,.
\end{equation}
For $\left|\delta_\alpha\right|<0.1$\,, as considered here, this would
lead to effects well below the theoretical accuracy quoted above and
therefore this effect was neglected and the variation of the $n+p \to
d + \gamma$ cross section with $\alpha$ is assumed to be entirely
given by Eq.~(\ref{eq:npdgvar})\,.

\subsection{\label{sec:Coulombenergies}Coulomb energies}

A variation of the value of the fine-structure constant $\alpha$
implies a variation of the nuclear binding energies and hence a
variation of the $Q$-values of the reactions, which in turn leads to a
variation of the cross sections and the corresponding rates.
Therefore the present study requires an estimate of the electromagnetic
contribution to the nuclear masses or equivalently to the nuclear
binding energies. A rough estimate is provided by the Coulomb term in
the Bethe-Weiszs\"acker formula (for a recent determination, see
\textit{e.g.}~\cite{Benzaid:2020prt} and references therein): 
\begin{equation}
  V^C_i = a_C\,\frac{Z_i\,(Z_i-1)}{A^\frac{1}{3}}\,,
  \qquad a_c \approx 0.64~\textnormal{MeV}\,,
\end{equation}
approximately accounting for the Coulomb repulsion by the protons in a
nucleus.  However, this formula is not very precise when applied to
the light nuclei relevant here.  We therefore prefer to use the
expectation values of the Coulomb interaction as determined from a
recent \textit{ab initio} calculation of light nuclear masses in the
framework of Nuclear Lattice Effective Field Theory
(NLEFT)~\cite{Elhatisari:2022qfr}\,, listed in Table~\ref{tab:VC}\,.
We also compare the calculated binding energies to the experimental
data as used here in order to give an impression of
the quality of the calculation.
\begin{table}[!htb]
  \caption{
    Binding energies $B$ (calculated ({\em cal}) and experimental
    ({\em exp}) values) and expectation values for the Coulomb interaction
    $V^C$ of light nuclei.\label{tab:VC}
  }
  \begin{ruledtabular}
    \begin{tabular}{@{}l%
      @{\extracolsep{\fill}}d{3.4}%
      @{\extracolsep{\fill}}d{3.4}%
      @{\extracolsep{\fill}}d{3.4}%
      @{}}
      nuclide
      & \mc{$V^C$[MeV]}%
        \footnote{%
        from~\cite{Elhatisari:2022qfr}.  The errors quoted in
        parentheses include all the statistic and systematic uncertainties.
        In case of $^2$H, the error is entirely given by the variation of the
        $np$ phase shifts at N3LO within their uncertainties.
        }
      & \mc{$B_{\textit{cal}}$[MeV]}%
        \footnotemark[1]
      & \mc{$B_{\textit{exp}}$[MeV]}%
        \footnote{%
        from~\cite{Huang:2021nwk}, as used in the present work.
        }
      \\
      \colrule
      ${}^{2}$H  & 0.0       &   2.215(150)    &   2.225 \\
      ${}^{3}$H  & 0.0       &   8.35(22) &   8.482 \\
      ${}^{3}$He & 0.688(1)  &   7.64(14) &   7.718 \\
      ${}^{4}$He & 0.759(0)  &  28.24(16) &  28.296 \\
      ${}^{6}$Li & 1.574(2)  &  32.82(12) &  31.994 \\
      ${}^{7}$Li & 1.599(2)  &  39.61(13) &  39.245 \\
      ${}^{8}$Li & 1.649%
                   \footnote{%
                   extrapolated from a least-squares fit to the other data with
                   $V^C(N,Z) = \sum_{k=0}^2\sum_{l=0}^1 c_{kl}(N-Z)^l(N+Z)^k$\,.
                   where
                   $c_{00}  =  0.653$\,,
                   $c_{10}  = -0.232$\,, 
                   $c_{20}  =  0.065$\,,
                   $c_{01}  =  0.060$\,,
                   $c_{11}  = -0.060$\,,
                   $c_{21}  = -0.003$ [in MeV]\,.
                   }
      &&  41.278 \\
      ${}^{7}$Be & 2.722\footnotemark[3]   && 37.600 \\ 
      ${}^{9}$Be & 2.951(4)  &  57.59(29) &  58.164 \\
      ${}^{8}$B  & 4.212\footnotemark[3]   &&  37.737 \\
      ${}^{10}$B & 4.453(8)  &  64.46(59) &  64.750 \\
      ${}^{11}$B & 4.962(2)  &  75.38(42) &  76.204 \\  
      ${}^{12}$B & 4.852\footnotemark[3]   &&  79.574 \\
      ${}^{11}$C & 6.933\footnotemark[3]   &&  73.440 \\
      ${}^{12}$C & 7.144(16) &  92.36(64) &  92.161 \\
      ${}^{13}$C & 7.151(7)  &  97.07(52) &  97.107 \\
      ${}^{14}$C & 7.317(7)  & 104.87(69) & 105.284 \\
      ${}^{12}$N & 9.483\footnotemark[3]   &&  74.040 \\
      ${}^{13}$N & 9.854\footnotemark[3]   &&  94.104 \\
      ${}^{14}$N & 10.354(4) & 106.25(94) & 104.657 \\
      ${}^{15}$N & 10.054(2) & 115.29(37) & 115.491 \\
      ${}^{14}$O & 12.977\footnotemark[3]  &&  98.730 \\
      ${}^{15}$O & 13.320\footnotemark[3]  && 111.953 \\
      ${}^{16}$O & 13.412(10)& 129.99(38) & 127.617 \\
    \end{tabular}
\end{ruledtabular}
\end{table}

\section{\label{sec:crearates}Calculation of the reaction rates}

For the 18 leading nuclear reactions in the BBN network,
\textit{viz.} the radiative capture reactions
\begin{equation}
  \label{eq:radcapreac}
  \begin{array}[c]{rclcrcl}
    n + p &\!\!\to\!\!\!& d + \gamma\,,
    &\quad&
            d + p &\!\!\to\!\!\!& \nuc{3}{He} + \gamma\,,
    \\
    p + \nuc{3}{H} &\!\!\to\!\!\!& \nuc{4}{He} + \gamma
    &\quad&
            d + \nuc{4}{He} &\!\!\to\!\!\!& \nuc{6}{Li} + \gamma\,,
    \\
    p + \nuc{6}{Li} &\!\!\to\!\!\!& \nuc{7}{Be} + \gamma\,,
    &\quad&
            \nuc{3}{H} + \nuc{4}{He} &\!\!\to\!\!\!& \nuc{7}{Li} + \gamma\,,
    \\
    \nuc{3}{He} + \nuc{4}{He} &\!\!\to\!\!\!& \nuc{7}{Be} + \gamma\,,
    &\quad& &&
  \end{array}
\end{equation}
the charged particle reactions
\begin{equation}
  \label{eq:chargedreac}
  \begin{array}[c]{rclcrcl} 
    d\!+\!d &\!\!\to\!\!\!& \nuc{3}{H}\!+\!p\,,
    &\quad&
            d\!+\!d &\!\!\to\!\!\!& \nuc{3}{He}\!+\!n\,,
    \\
    d\!+\!\nuc{3}{H} &\!\!\to\!\!\!& \nuc{4}{He}\!+\!n\,,
    &\quad&
    d\!+\!\nuc{3}{He} &\!\!\to\!\!\!& \nuc{4}{He}\!+\!p\,,
    \\
    p\!+\!\nuc{6}{Li} &\!\!\to\!\!\!& \nuc{3}{He}\!+\!\nuc{4}{He}\,,
    &\quad&
            p\!+\!\nuc{7}{Li} &\!\!\to\!\!\!& \nuc{4}{He}\!+\!\nuc{4}{He}\,,
    \\
    d\!+\!\nuc{7}{Li} &\!\!\to\!\!\!& \nuc{4}{He} \!+\!\nuc{4}{He}\!+\!n\,,
    &\quad&                      
            d\!+\!\nuc{7}{Be} &\!\!\to\!\!\!& \nuc{4}{He}\!+\!\nuc{4}{He}\!+\!p\,,
  \end{array}
\end{equation}
and the neutron-induced reactions
\begin{equation}
  \label{eq:nindreac}
  \begin{array}[c]{rclcrcl} 
    n\!+\!\nuc{3}{He} &\!\!\to\!\!\!& \nuc{3}{H}\!+\!p\,,
    &\quad&
            n\!+\!\nuc{7}{Be} &\!\!\to\!\!\!& \nuc{7}{Li}\!+\!p\,,
    \\
    n\!+\!\nuc{7}{Be} &\!\!\to\!\!\!& \nuc{4}{He}\!+\!\nuc{4}{He}\,,
                          &&
  \end{array}
\end{equation}
the rates and their variations with $\alpha$ are calculated by a
numerical integration of Eq.~(\ref{eq:ratevar}) and tabulated for 60
temperatures in the range $0.001 \le T_9 = T/[10^9~\textnormal{K}] \le
10.0$.  These values were then used via a cubic spline interpolation
in the four publicly available BBN codes as outlined in
Sect.~\ref{sec:response}. The resulting rates and their variations
with $\alpha=\alpha_0\,(1+\delta_\alpha)$ in the range $\delta_\alpha
\in [-0.1,0.1]$ are displayed in Fig.~\ref{fig:ratealpha} in
Sect.~\ref{sec:approx}. To this end, we made new fits to the cross
sections (or equivalently of the corresponding astrophysical
$S$-factors) for the reactions listed above. The parameterizations can
be found in Appendix~\ref{app:params}.  In addition in
Fig.~\ref{fig:rates_comp_all} the resulting reaction rates for
$\alpha=\alpha_0$ are compared to the rates implemented in the
original versions of the four programmes considered here.
\begin{figure*}[!htb] 
  \includegraphics[width=\textwidth, trim=25 230 70 70, clip]{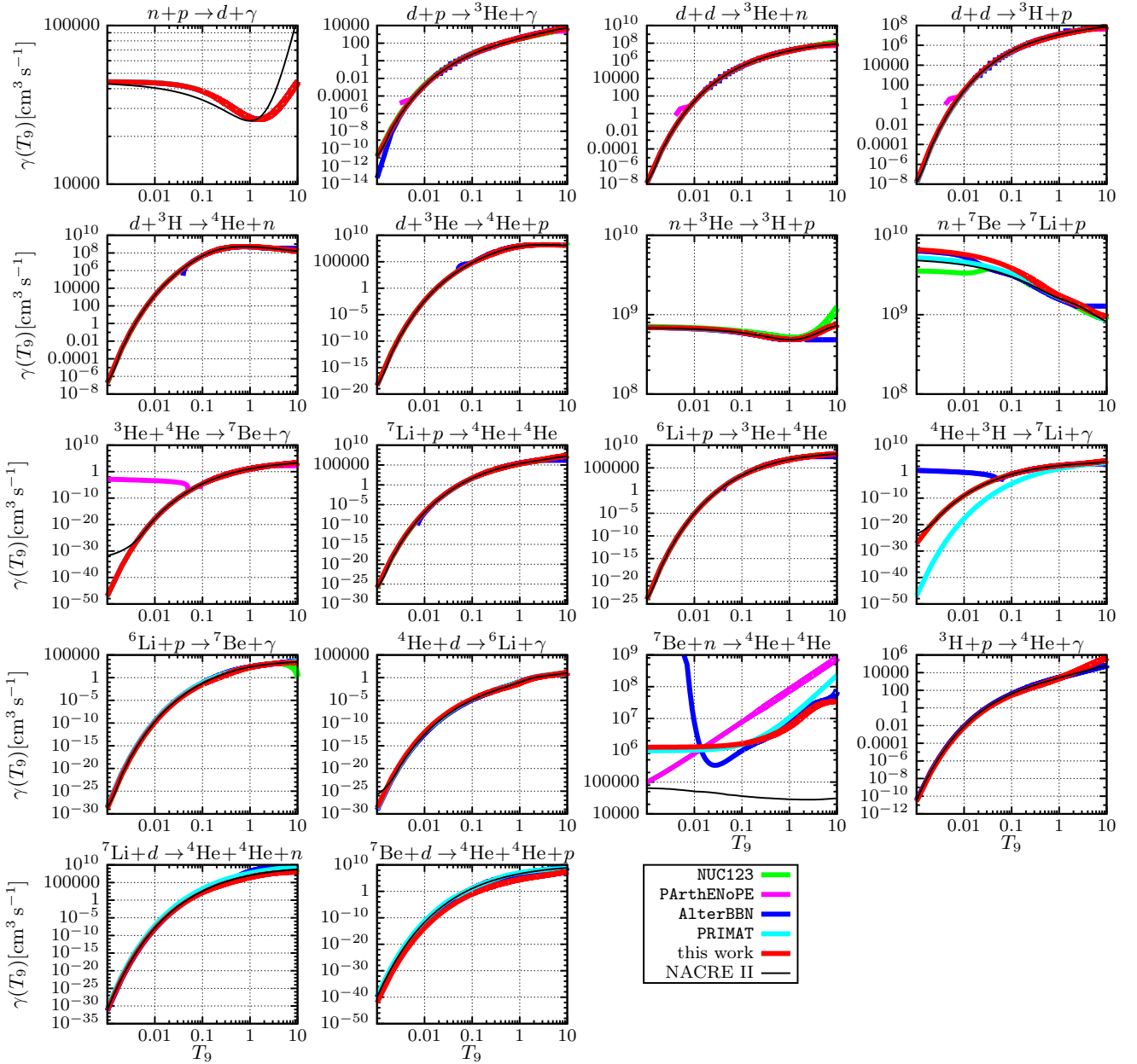}
  \caption{\label{fig:rates_comp_all}
    Reaction rates $\gamma(T_9)$ for 18 leading nuclear reactions in the BBN
    network, where $T_9 = T/[10^9~\textnormal{K}]$\,. The rates resulting
    from the new parameterizations of the $S$-factors in Appendix~\ref{app:params}
    are represented by solid red curves (color online). The rates in the
    original version of the programmes are given by green curves for
    \texttt{NUC123}~\cite{Kawano:1992ua}, magenta curves for
    \texttt{PArthENoPE}~\cite{Gariazzo:2021iiu}, blue curves for
    \texttt{AlterBBN}~\cite{Arbey:2018zfh} and cyan curves for the
    \texttt{PRIMAT}~\cite{Pitrou:2018cgg} code. Also shown as a thin black
    curve is the result from the NACRE II database,
    see~\cite{Xu:2012uw}\,.
  } 
\end{figure*}
In Fig.~\ref{fig:rates_comp_all} we also display the rates
obtained with the NACRE II database, see~\cite{Xu:2012uw}, which
served as a further check on our calculated reaction rates at $\alpha
= \alpha_0$\,.  The rates of all other reactions were taken as in the
original implementation of the codes and the variation of the rates
with $\alpha$ was calculated as discussed in Sect.~\ref{sec:approx}.

The variation of the $\beta$-decay rates according to
Eq.~(\ref{eq:lambdabeta}) was implemented directly in the various
codes. In Fig.~\ref{fig:alphabetaRQD} it is shown how the
$\beta$-rates at low temperature ({\em i.e.} $T \ll T_9$)
change by a variation of $\alpha =
\alpha_0\,(1+\delta_\alpha)$. In particular the rates of the tritium
decay and the $\nuc{14}{C}$-decay strongly depend on the value of
$\delta_\alpha$, the effect of (relatively large) changes in the
(relatively small) $Q$-values due to changes in the Coulomb
contribution to the binding energies being dominant.

\begin{figure}[!htb]
  \centering
  \includegraphics[width=\columnwidth, clip]{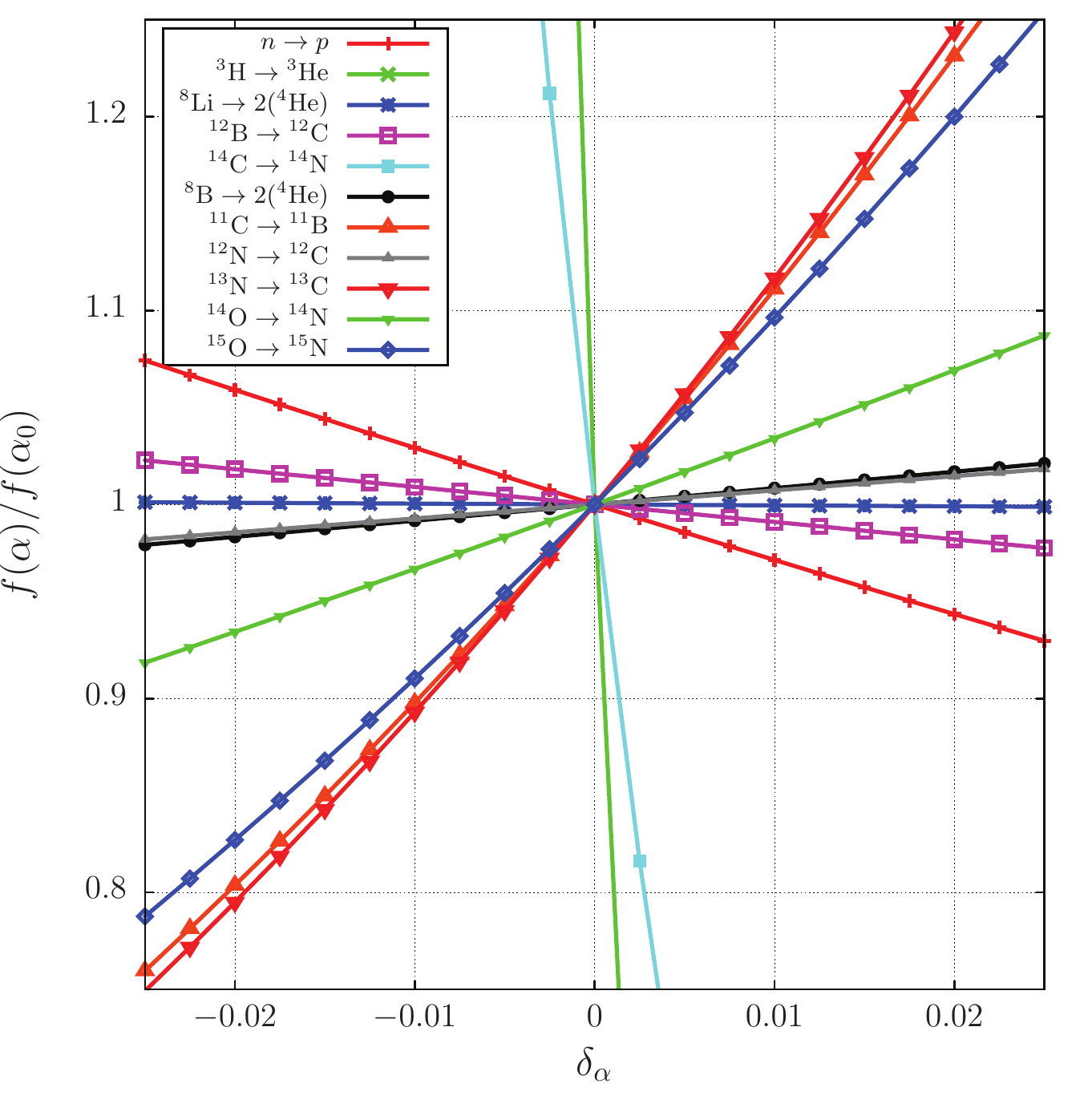}
  \caption{\label{fig:alphabetaRQD}
    Fractional variation of the $\beta$-rates at low temperature with
    $f(\alpha)/f(\alpha_0)$ calculated by Eq.~(\ref{eq:lambdabeta}).
  } 
\end{figure}

As already touched upon in section~\ref{sec:weaknp} the variation of
the weak $n \leftrightarrow p$ rates with $\alpha$ is dominated by the
variation of proton-neutron mass difference with $\alpha$ and is
strongly temperature dependent in the early phase of BBN. In the
default version of the Kawano code
\texttt{NUC123}~\cite{Kawano:1992ua} this temperature dependence is
parameterized as outlined in Appendix F of Ref.~\cite{Kawano:1992ua},
but a numerical integration along Eq.~(\ref{eq:Gnp}) can be enforced
and was in fact used to implement the $\alpha$-dependence of these
rates.  The \texttt{PArth\-ENoPE}
code~\cite{Pisanti:2007hk,Consiglio:2017pot,Gariazzo:2021iiu} contains
a slightly more sophisticated parameterization, see \textit{e.g.}
Appendix C of Ref.~\cite{Serpico:2004gx}, accounting also for some
higher order corrections.  Here we used the $\alpha$-dependence of the
$n \leftrightarrow p$ rates as illustrated in Fig.~\ref{fig:nprate} as
a factor multiplying the parameterized rate. In the \texttt{AlterBBN}
code~\cite{Arbey:2011nf, Arbey:2018zfh} the temperature dependence of
the weak $n \leftrightarrow p$ rates was already determined
numerically as in Eq.~(\ref{eq:Gnp}) and the $\alpha$ dependence can
be accounted for by an appropriate variation of $\Delta m$\,.  In this
code also the Coulomb correction, see Eq.~(\ref{eq:lambdabeta}), was
included in the integrand of Eq.~(\ref{eq:Gnp}), but this was found to
have no significant impact on the final abundances to be discussed
below in section~\ref{sec:results}.  The
\texttt{PRIMAT}~\cite{Pitrou:2018cgg} implementation offers the
possibility to study the $\alpha$-dependence of the weak $n
\leftrightarrow p$ reactions in all detail including all the higher
order electromagnetic corrections mentioned in
section~\ref{sec:weaknp}\,. In fact this code was used to verify that
the variation of the rates through the variation of $\Delta m$ with
$\alpha$ as discussed in section~\ref{sec:weaknp} is indeed the
dominant effect.  Indeed, ignoring the $\alpha$ dependence in the
higher order corrections implemented in \texttt{PRIMAT} led to
response coefficients that differ at most by 0.5\% from the values
listed in Table~\ref{tab:rme} below.  Accordingly, in spite of the
fact that the $n \leftrightarrow p$ reactions are treated at various
levels of sophistication, the resulting primordial abundances and
their variation with $\alpha$, to be discussed in
section~\ref{sec:results}, were found to be rather consistent.

\section{\label{sec:response}The BBN response matrix}

We estimated the linear dependence of the primordial abundances $Y_n$
on small changes in the value of the fine-structure constant $\alpha =
\alpha_0\,(1+\delta_\alpha)$ by calculating the abundance of the
nuclide $n$, with
\begin{equation}
  n \in \lbrace
  {^2}\textnormal{H}\,,
  {^3}\textnormal{H}+{^3}\textnormal{He}\,,
  {^4}\textnormal{He}\,,
  {^6}\textnormal{Li}\,,
  {^7}\textnormal{Li}+{^7}\textnormal{Be}
  \rbrace
\end{equation}
\textit{i.e.} $Y_n(\alpha_0\,(1+\delta_\alpha))$, for fractional
changes $\delta_\alpha$ in the range $[-0.1,0.1]$ with the publicly
available codes for BBN, name\-ly a version of the Kawano code
\texttt{NUC123}~\cite{Kawano:1992ua} (in \texttt{FORTRAN}), two more
modern implementations based on this, namely
\texttt{PArth\-ENoPE}~\cite{Pisanti:2007hk,Consiglio:2017pot,Gariazzo:2021iiu}
(\texttt{in FORTRAN}) and \texttt{AlterBBN}~\cite{Arbey:2011nf,
Arbey:2018zfh} (in \texttt{C}) as well as an implementation as a
\texttt{mathematica}-note{\-}book,
\texttt{PRIMAT}~\cite{Pitrou:2018cgg}\,.  To this end we performed
least-squares fits of a quadratic polynomial to the abundances:
\begin{equation}
  \label{eq:quadfit}
  P_k(\delta_\alpha) = c_0\left(1+c_1\,\delta_\alpha^{} + c_2\,\delta_\alpha^2\right)\,,
\end{equation}
such that
\begin{equation}
  \diff{}{c_j}\left|Y_n{/Y_H}(\alpha_0\,(1+\delta_\alpha))-P_k(\delta_\alpha)\right|^2 = 0\,,\quad j=0,1,2\,.
\end{equation}
Then
\begin{equation}
  \diff{\log{{(Y_n/Y_H)}}}{\log{\alpha}}
  \approx
  c_1
\end{equation}
will be called an element of the linear nuclear BBN response
matrix. It represents the dimensionless fractional chan\-ge in the
primordial abundance ratio $Y_n{/Y_H}$ due to a fractional change
$\alpha$ in linear approximation.  Deviations from a linear response
are then given by the coefficient $c_2$\,.

\section{\label{sec:results}Results and discussion}

In most of what follows we shall use $\eta=6.14\cdot 10^{-10}$ from
Ref.~\cite{Workman:2022ynf} as the nominal baryon-to-photon density
ratio while varying $\alpha$. The programs were modified as indicated
in Sect.~4 of Ref.~\cite{Meissner:2022dhg} and the rates for the most
relevant reactions listed in Sect.~\ref{sec:crearates}, resulting from
the new fits of the cross sections presented in
Appendix~\ref{app:params}, were used in all programmes.

\begin{table}[hbt]
  \begin{ruledtabular}
      \caption{\label{tab:finab}%
        Final abundances as number ratios $Y_n/Y_H$ (for
        ${}^4\textnormal{He}$ the mass ratio $Y_p$) calculated with the
        modified versions of the codes. The value of the baryon-to-photon
        ratio and the nominal value of the neutron lifetime are $\eta = 6.14
        \cdot 10^{-10}$ and $\tau_n = 879.4\,\textnormal{s}$, respectively.
        For comparison also the values previously obtained in 
        Ref.~\cite{Meissner:2022dhg} are listed.      }
      \begin{tabular}{@{}l%
        @{\extracolsep{\fill}}d{3.4}%
        @{\extracolsep{\fill}}d{3.4}%
        @{\extracolsep{\fill}}d{3.4}%
        @{\extracolsep{\fill}}d{3.4}%
        @{\extracolsep{\fill}}d{3.4}%
        @{}}
        \texttt{code}
        & \mc{$\nuc{2}{H}$}
        & \mc{$\nuc{3}{H}\!\!+\!\!\nuc{3}{He}$}
        & \mc{$Y_{p}$}
        & \mc{$\nuc{6}{Li}$}
        & \mc{$\nuc{7}{Li}\!\!+\!\!\nuc{7}{Be}$}
        \\[-0.25ex]
        & \mc{$\times 10^5$}
        & \mc{$\times 10^5$}
        & \mc{}
        & \mc{$\times 10^{14}$}
        & \mc{$\times 10^{10}$}
        \\
        \colrule
        \texttt{NUC123}        & 2.501 & 1.139 & 0.246 & 1.809 & 5.172 \\
        \cite{Meissner:2022dhg}& 2.550 & 1.040 & 0.247 & 1.101 & 4.577 \\
        \texttt{PArthENoPE}    & 2.569 & 1.147 & 0.247 & 1.820 & 5.017 \\
        \cite{Meissner:2022dhg}& 2.511 & 1.032 & 0.247 & 1.091 & 4.672 \\
        \texttt{AlterBBN}      & 2.585 & 1.153 & 0.248 & 1.904 & 4.993 \\
        \cite{Meissner:2022dhg}& 2.445 & 1.031 & 0.247 & 1.078 & 5.425 \\
        \texttt{PRIMAT}        & 2.563 & 1.149 & 0.247 & 1.862 & 5.033 \\
        \cite{Meissner:2022dhg}& 2.471 & 1.044 & 0.247 & 1.198 & 5.413 \\
        \colrule
        PDG~\cite{Workman:2022ynf}  & 2.547 &       & 0.245 &       & 1.6 \\
        $\qquad\pm$  & 0.025 &       & 0.003 &       & 0.3 \\  
      \end{tabular}
  \end{ruledtabular}
\end{table}   
The resulting nominal (\textit{i.e.} at $\alpha=\alpha_0$) abundances
at the end of the BBN epoch in terms of the number ratios
$Y_{{}^2\textnormal{H}}/Y_\textnormal{H}$\,,
$Y_{{}^3\textnormal{H}+{}^3\textnormal{He}}/Y_\textnormal{H}$\,,
$Y_{{}^6\textnormal{Li}}/Y_\textnormal{H}$\,,
$Y_{{}^7\textnormal{Li}+{}^7\textnormal{Be}}/Y_\textnormal{H}$\,,
and the mass ratio for ${}^4\textnormal{He}$ are compared to the
values quoted in Ref.~\cite{Meissner:2022dhg} and experimental data in
Table~\ref{tab:finab}\,.  Although the mass ratio for $\nuc{4}{He}$
and, to a minor extend, the number ratio for deuterium did not
change significantly with respect to the values obtained in
Ref~\cite{Meissner:2022dhg}, the $\nuc{3}{H}+\nuc{3}{He}$ number ratio
increased by approximately $10\%$\,, the $\nuc{6}{Li}$ number ratio
was found to be larger by about $70-80\%$\,, while the
$\nuc{7}{Li}+\nuc{7}{Be}$ number ratio is still too large by a factor
of three, a phenomenon known as the lithium-problem, which is thus
unsolved even using the updated cross sections used here. As stated
previously in~\cite{Meissner:2022dhg}, in spite of this unresolved
issue in BBN the consistency of the cosmic microwave background
observations with the determined abundances of deuterium and helium is
considered to be a non-trivial success. Accordingly, we think that
this issue is no obstacle for the study presented here.

The elements of the response matrix were then determined by a
polynomial fit, as explained above in Sect.~\ref{sec:response} for the
abundances relative to the hydrogen abundance, namely
$Y_{{}^2\textnormal{H}}/Y_\textnormal{H}$\,,
$Y_{{}^3\textnormal{H}+{}^3\textnormal{He}}/Y_\textnormal{H}$\,,
$Y_{{}^6\textnormal{Li}}/Y_\textnormal{H}$\,,
$Y_{{}^7\textnormal{Li}+{}^7\textnormal{Be}}/Y_\textnormal{H}$\,, 
and the mass ratio for ${}^4\textnormal{He}$\,. 

The dependence of these ratios on the value of the fine structure
constant $\alpha = \alpha_0\,(1+\delta_\alpha)$ is displayed in
Fig.~\ref{fig:fabund} for $\delta_\alpha \in [-0.1,0.1]$.
\begin{figure*}[!htb]
  \centering
  \includegraphics[width=0.97\textwidth, trim=105 325 65 50, clip]{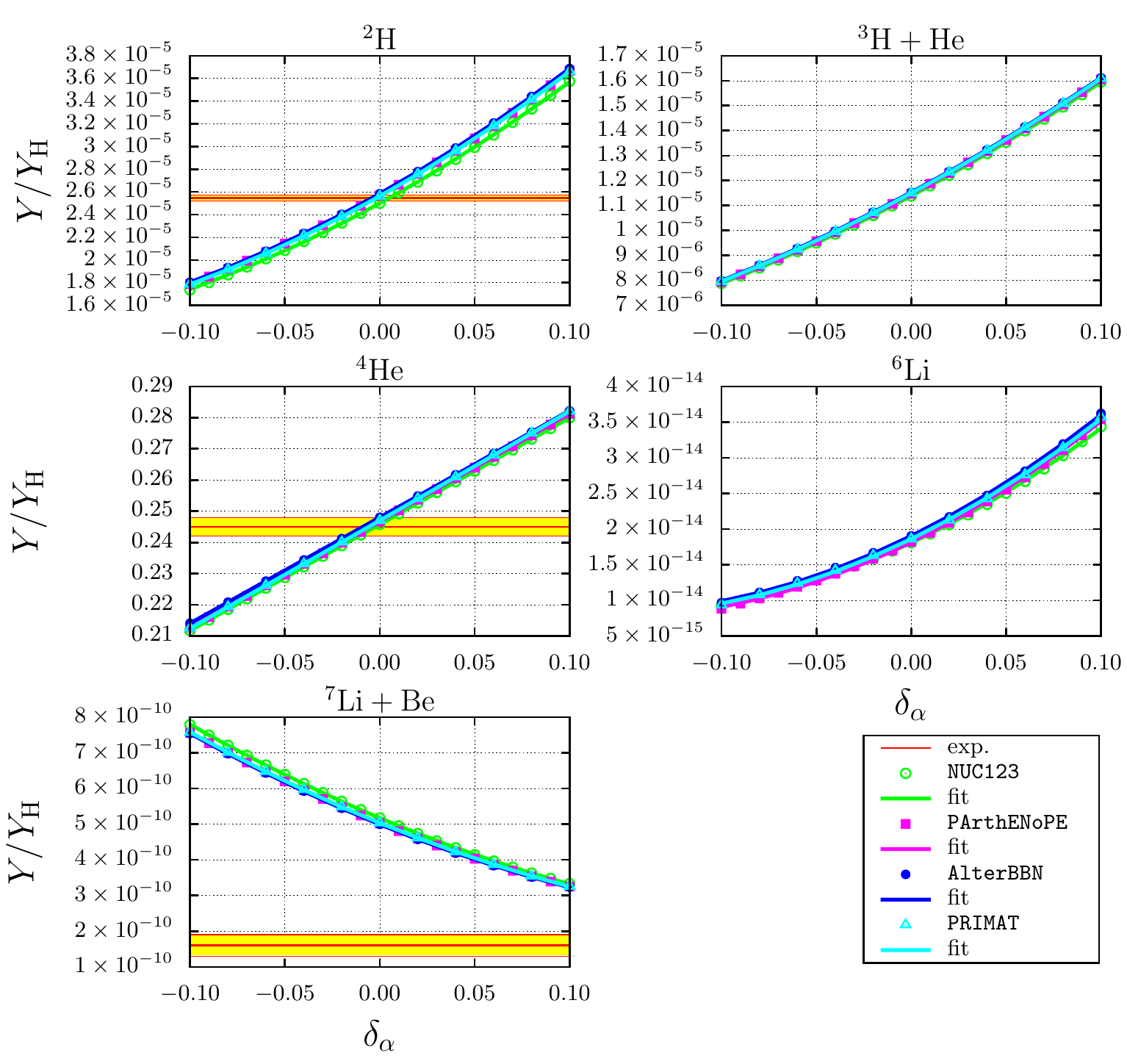}
  \caption{  \label{fig:fabund}
    Variation of the abundance ratios $Y_n/Y_H$ with a variation of $\alpha =
    \alpha_0\,(1+\delta_\alpha)$ for $\delta_\alpha \in [-0.1,0.1]$ obtained with the codes:
    \texttt{NUC123}~\cite{Kawano:1992ua},
    \texttt{AlterBBN}~\cite{Arbey:2018zfh},
    \texttt{PArthENoPE}~\cite{Gariazzo:2021iiu},
    \texttt{PRIMAT}~\cite{Pitrou:2018cgg}\,. 
    Here, we use
    $\eta=6.14 \cdot 10^{-10}$
    and
    $\tau_n = 879.4\,\textnormal{s}$\,. 
    Also shown are the solid curves obtained by the fits according
    to Eq.~(\ref{eq:quadfit}) with the parameters listed in
    Table~\ref{tab:rme}.  The experimental values cited in
    PDG~\cite{Workman:2022ynf} (thick red lines) are indicated by
    yellow-highlighted regions (color online) representing the $1\sigma$
    limits by red lines.
  }
\end{figure*}
Indeed the variation of the abundance ratios is found to be very
similar for all four publicly available codes, in spite of the fact
that these codes differ in details, such as the number of reactions in
the BBN network or the manner in which the rate equations are solved
numerically.  Note, however, that in the present study the rates
calculated for the major reactions listed in Sect.~\ref{sec:crearates}
and their variation with $\alpha$ are the same.

Of course this then also applies to the values for the resulting
response matrix elements.  The response matrix elements
$\partial\log{(Y_n/Y_H)}/\partial\log{\alpha} = c_1$ and the
coefficients of the quad{\-}ratic term in Eq.~(\ref{eq:quadfit}) are
given and compared to some results from the literature in
Table~\ref{tab:rme}.  Note that with the exception of $\nuc{6}{Li}$,
we have $|c_2| \simeq |c_1|$, so that due to the smallness of
$\alpha$, the second order contribution to the response is of minor
importance.  All programs were run with the full network implemented
in the original version codes.  We checked that if we run the programs
with a smaller network the results listed in
Tables~\ref{tab:finab},\ref{tab:rme} change only in the last digit and
therefore conclude that the approximation, see Sect.~\ref{sec:approx},
we made for rate changes in the reactions beyond the reactions listed
in Eqs.~(\ref{eq:radcapreac}--\ref{eq:nindreac}) are without any
effect for the present investigation.

\begin{table*}[hbt]
  \caption{\label{tab:rme}%
    BBN response matrix $c_1
    = \partial\log{(Y_n/Y_H)}/\partial\log{\alpha}$ and the coefficients $c_2$
    of the quadratic term in Eq.~(\ref{eq:quadfit}) at $\eta=6.14 \cdot 10^{-10}$
    and $\tau_n = 879.4\,\textnormal{s}$\,. $Y_n/Y_H$ are the number ratios of
    the abundances relative to hydrogen; $Y_p$ is conventionally the
    ${}^4$He/H mass ratio. The results obtained with the four BBN codes
    \texttt{NUC123}~\cite{Kawano:1992ua},
    \texttt{PArthENoPE}~\cite{Gariazzo:2021iiu},
    \texttt{AlterBBN}~\cite{Arbey:2018zfh},
    \texttt{PRIMAT}~\cite{Pitrou:2018cgg} are given in four subsequent rows
    and compared to earlier results from Refs.~\cite{Dent:2007zu,Nollett:2002da,Bergstrom:1999wm}.
  }
  \begin{ruledtabular}
    \begin{tabular}{@{}l%
        @{\extracolsep{2.0em}}d{3.4}%
        @{\extracolsep{0.4em}}d{3.4}%
        @{\extracolsep{2.0em}}d{3.4}%
        @{\extracolsep{0.4em}}d{3.4}%
        @{\extracolsep{2.0em}}d{3.4}%
        @{\extracolsep{0.4em}}d{3.4}%
        @{\extracolsep{2.0em}}d{3.4}%
        @{\extracolsep{0.4em}}d{3.4}%
        @{\extracolsep{2.0em}}d{3.4}%
        @{\extracolsep{0.4em}}d{3.4}%
        @{}} 
      \texttt{code}
      & \multicolumn{2}{c}{$\nuc{2}{H}$}
      & \multicolumn{2}{c}{$\nuc{3}{H}\!\!+\!\!\nuc{3}{He}$}
      & \multicolumn{2}{c}{$Y_{p}$}
      & \multicolumn{2}{c}{$\nuc{6}{Li}$}
      & \multicolumn{2}{c}{$\nuc{7}{Li}\!\!+\!\!\nuc{7}{Be}$}
      \\
      \cline{2-3}\cline{4-5}\cline{6-7}\cline{8-9}\cline{10-11}
      & \mc{$c_1$} & \mc{$c_2$}
      & \mc{$c_1$} & \mc{$c_2$}
      & \mc{$c_1$} & \mc{$c_2$}
      & \mc{$c_1$} & \mc{$c_2$}
      & \mc{$c_1$} & \mc{$c_2$}
      \\
      \colrule
      \texttt{NUC123}     &  3.655 & 6.228 & 3.540 & 4.625 & 1.387 & 0.016 & 6.830 & 20.412 & -4.325 & 7.480 \\
      \texttt{PArthENoPE} &  3.635 & 6.182 & 3.533 & 4.577 & 1.389 & 0.065 & 7.159 & 21.482 & -4.308 & 7.715 \\
      \texttt{AlterBBN}   &  3.644 & 6.188 & 3.526 & 4.568 & 1.373 & 0.049 & 6.857 & 20.499 & -4.322 & 7.865 \\
      \texttt{PRIMAT}     &  3.658 & 6.264 & 3.534 & 4.595 & 1.408 & 0.081 & 6.953 & 20.828 & -4.302 & 7.563 \\
      \colrule
      \cite{Dent:2007zu}  &  3.6    &        & 0.95   &        & 1.9    &        & 6.6    &         & -11     &        \\
      \cite{Nollett:2002da}\footnote{%
      Extracted from Fig.2 of~\cite{Nollett:2002da} for $\eta=5.6\cdot 10^{-10}$
      after digitizing the data.
      }
      & 3.99   & 5.99   & 1.04   &-2.67   &        &        &        &         & -9.30   & 25.7   \\
      \cite{Bergstrom:1999wm}\footnote{%
      Extracted from Fig.4 of~\cite{Bergstrom:1999wm} for $\eta=5\cdot10^{-10}$
      after digitizing the data.
      } 
      & 5.13 & 9.91  & 0.78   &-1.96   & 1.96   & -0.73   &        &         & -13.6   & 83.1     \\ 
    \end{tabular}
  \end{ruledtabular}
\end{table*}  
Apart from the values of $c_1$ for $\nuc{2}{H} (\approx{3.6})$ and for
$\nuc{6}{Li} (\approx{6.8})$ the values obtained in the present study,
although consistent among each other, differ appreciably from the
values obtained in Refs.~\cite{Dent:2007zu, Nollett:2002da,
Bergstrom:1999wm}. In particular in the present calculations the
linear response for $\nuc{3}{H}+\nuc{3}{He}$ is much larger while the
linear response for $\nuc{7}{Li}+\nuc{7}{Be}$ is appreciably smaller
in magnitude, although there seems to be at least a consensus
concerning the sign.

In order to clarify this issue, we shall discuss in some detail the
relevance of the various factors that reflect the $\alpha$-dependence
of the nuclear rates: 
\begin{itemize}
\item
  First of all we list in Table~\ref{tab:rmebeta} the linear response
  of the BBN abundances to a variation of $\alpha$ in the $\beta$-decay
  rates only.
  \begin{table}[!htb]
    \caption{%
      BBN response matrix $c_1 = \partial\log( Y_n/Y_H)/\partial\log\alpha$
      accounting for the variation of the $\beta$-decay rates only. See
      also the caption of Table~\ref{tab:rme}\,.\label{tab:rmebeta}
    }
    \begin{ruledtabular}
      \begin{tabular}[c]{@{}l%
        @{\extracolsep{\fill}}d{3.4}%
        @{\extracolsep{\fill}}d{3.4}%
        @{\extracolsep{\fill}}d{3.4}%
        @{\extracolsep{\fill}}d{3.4}%
        @{\extracolsep{\fill}}d{3.4}%
        @{}}
        \texttt{code}
        & \mc{$\nuc{2}{H}$}
        & \mc{$\nuc{3}{H}+\nuc{3}{He}$}
        & \mc{$Y_p$}
        & \mc{$\nuc{6}{Li}$}
        & \mc{$\nuc{7}{Li} + \nuc{7}{Be}$}
        \\
        \colrule
        \texttt{NUC123}     & 0.827 & 0.250 & 1.403 & 2.651 & 0.475 \\
        \texttt{PArthENoPE} & 0.832 & 0.255 & 1.406 & 2.663 & 0.479 \\
        \texttt{AlterBBN}   & 0.829 & 0.255 & 1.390 & 2.632 & 0.462 \\
        \texttt{PRIMAT}     & 0.845 & 0.260 & 1.425 & 2.701 & 0.483 \\
      \end{tabular}
    \end{ruledtabular}
  \end{table}
\item
  In Table~\ref{tab:rmenuclear} we display the linear response
  of the BBN abundances to a variation of the nuclear reaction rates.
  \begin{table}[!htb]
    \caption{%
      BBN response matrix $c_1 = \partial\log( Y_n/Y_H)/\partial\log\alpha$
      accounting for the variation of the nuclear rates only, but also including
      the variation of the binding energies and thus of the $Q$-values of
      the reactions. See also the caption of
      Table~\ref{tab:rme}\,.\label{tab:rmenuclear}
      }
    \begin{ruledtabular}
      \begin{tabular}[c]{@{}l%
        @{\extracolsep{\fill}}d{3.4}%
        @{\extracolsep{\fill}}d{3.4}%
        @{\extracolsep{\fill}}d{3.4}%
        @{\extracolsep{\fill}}d{3.4}%
        @{\extracolsep{\fill}}d{3.4}%
        @{}}
        \texttt{code}
        & \mc{$\nuc{2}{H}$}
        & \mc{$\nuc{3}{H}+\nuc{3}{He}$}
        & \mc{$Y_p$}
        & \mc{$\nuc{6}{Li}$}
        & \mc{$\nuc{7}{Li} + \nuc{7}{Be}$}
        \\
        \colrule
        \texttt{NUC123}     & 2.818 & 3.271 & -0.017 & 4.005 & -5.192 \\
        \texttt{PArthENoPE} & 2.795 & 3.261 & -0.017 & 4.315 & -5.152 \\
        \texttt{AlterBBN}   & 2.806 & 3.254 & -0.017 & 4.037 & -5.153 \\
        \texttt{PRIMAT}     & 2.803 & 3.257 & -0.017 & 4.059 & -5.164 \\
      \end{tabular}
    \end{ruledtabular}
  \end{table}
  The relevance of the variation of the binding energies with $\alpha$
  may be appreciated by the linear response due to changes in $\alpha$
  accounting for the effects due to the Coulomb penetration factors
  only, \textit{i.e.} without accounting for changes in the binding
  energies, listed in Table~\ref{tab:rmenuclrat}\,.
  Here we also compared our results to the results presented in Table~I 
  of Ref.~\cite{Dent:2007zu} for the dependence of the abundances on
  the nuclear rate variation with $\alpha$, that thus differ from our
  results significantly for $c_1(\nuc{3}{H}+\nuc{3}{He})$ and
  $c_1(\nuc{7}{Li}+\nuc{7}{Be})$\,, our results being larger in magnitude
  for the former and smaller for the latter.
  \begin{table}[!htb]
    \caption{%
      BBN response matrix $c_1 = \partial\log( Y_n/Y_H)/\partial\log\alpha$
      accounting for the variation of the nuclear rates only, but excluding 
      the variation of the binding energies. Also see caption to
      Table~\ref{tab:rme}\,.\label{tab:rmenuclrat}
    }
    \begin{ruledtabular}
      \begin{tabular}[c]{@{}l%
        @{\extracolsep{\fill}}d{3.4}%
        @{\extracolsep{\fill}}d{3.4}%
        @{\extracolsep{\fill}}d{3.4}%
        @{\extracolsep{\fill}}d{3.4}%
        @{\extracolsep{\fill}}d{3.4}%
        @{}}
        \texttt{code}
        & \mc{$\nuc{2}{H}$}
        & \mc{$\nuc{3}{H}+\nuc{3}{He}$}
        & \mc{$Y_p$}
        & \mc{$\nuc{6}{Li}$}
        & \mc{$\nuc{7}{Li} + \nuc{7}{Be}$}
        \\
        \colrule
        \texttt{NUC123}     & 2.619 & 3.559 & -0.016 & 5.561 & -2.059 \\
        \texttt{PArthENoPE} & 2.599 & 3.550 & -0.017 & 5.866 & -2.005 \\
        \texttt{AlterBBN}   & 2.598 & 3.557 & -0.016 & 5.585 & -1.758 \\
        \texttt{PRIMAT}     & 2.595 & 3.562 & -0.017 & 5.610 & -1.769 \\
        \cite{Dent:2007zu}  & 2.3   & 0.79  &  0.00  & 4.6   & -8.1    \\
      \end{tabular}
    \end{ruledtabular}
  \end{table}
\end{itemize}
Indeed, if we substitute our values, as well as the results we
obtained in~\cite{Meissner:2022dhg} for the linear response of the
abundances on binding energies and the neutron life-time $\tau_n$ for
the values of the response matrix $C$ of Table~I in~\cite{Dent:2007zu}
and furthermore account for the smaller value
$\partial\log{\tau_n}/\partial\log\alpha\approx 2.90$, obtained via
Eq.~(\ref{eq:lambdabeta}) (instead of $3.86$ in~\cite{Dent:2007zu})
and the smaller value $\partial\log{Q_N}/\partial\log\alpha\approx
-0.45$ (instead of $-0.59$ in~\cite{Dent:2007zu}), due to the smaller
new value for $\Delta m_{\textnormal{\tiny{QED}}}$, and use our values
for the response of the binding energies
$\partial\log{B_i}/\partial\log\alpha$ that are smaller by about
$10\%$ in Table IV of~\cite{Dent:2007zu} we find approximately for the
linear responses
\begin{center}
  \begin{tabular}[c]{@{}l%
  @{\extracolsep{\fill}}d{3.4}%
  @{\extracolsep{\fill}}d{3.4}%
  @{\extracolsep{\fill}}d{3.4}%
  @{\extracolsep{\fill}}d{3.4}%
  @{\extracolsep{\fill}}d{3.4}%
  @{}}
    \colrule
    \\[-1.25ex]
  & \mc{$\nuc{2}{H}$}
  & \mc{$\nuc{3}{H}+\nuc{3}{He}$}
  & \mc{$Y_p$}
  & \mc{$\nuc{6}{Li}$}
  & \mc{$\nuc{7}{Li} + \nuc{7}{Be}$}
  \\[0.25ex]
  \colrule
  \\[-1.25ex]
  & 3.{7} & 3.5  & 1.4 & 7.0 & -4.4 \\[0.5ex] 
  \colrule
  \end{tabular}
\end{center}
rather close to our values for $c_1$ given in Table~\ref{tab:rme}.
Most of the effects listed above are, although significant, of minor
importance only, and accordingly the difference can be traced back to
the fact that our results for the variation of the rates with a
variation of $\alpha$ when ignoring the effects based on $Q$-value
changes, as listed in Table~\ref{tab:rmenuclrat} differ appreciably
from those of~\cite{Dent:2007zu}\,. Unfortunately in the latter
reference no results on the $\alpha$-dependence of the rates are
explicitly given. In appendix~A.2 of~\cite{Dent:2007zu} it is
mentioned that parameterizations of the $S$-factors were used, the
parameters determined by fitting the NETGEN rates as closely as
possible. In order to check our parameterizations of the nuclear rates
we compared our rates with results generated by the
NETGEN-tool~\cite{Xu:2012uw} in Fig.~\ref{fig:rates_comp_all} and
found that these are indeed compatible for all reactions, except for
the reaction $\nuc{7}{Be}+n \to \nuc{4}{He} + \nuc{4}{He}$, where the
NETGEN-tool merely uses the THALYS nuclear reaction
model~\cite{Goriely:2008zu}. We instead used data, see also
Fig.~\ref{fig:7Benaa} for our fit of the $S$-factor. Therefore the
difference must be due to the different way the Coulomb penetration
effects are treated. Note that, as emphasized in
Sect.~\ref{sec:penetration}, we did not rely on
temperature-independent penetration factors taken as a Gamow-factor,
but rather accounted for the penetration dependences in the
cross-section, which then leads to temperature-dependent changes in
the rates.

Our results also differ from the results in
Refs.~\cite{Nollett:2002da} and~\cite{Bergstrom:1999wm} published even
earlier. Concerning the treatment in~\cite{Nollett:2002da}, it is
noted that, although the authors present a detailed discussion of the
$\alpha$-dependence in the penetration factors, even accounting for
additional $\alpha$-dependent effects due to the peripheral nature of
some radiative capture reactions such as \textit{e.g.} the
$\nuc{3}{He}+\nuc{4}{He} \to \nuc{7}{Be} + \gamma$\,, an effect taken
into account also in the present treatment. Nevertheless, in contrast
to our treatment, $\alpha$-dependent effects seem to be treated merely
by temperature-independent factors multiplying the rates.  In
Ref.~\cite{Bergstrom:1999wm}, the changes in the reaction rates due to
changes in $\alpha$ were treated through approximate expressions based
on expansions of the $S$-factors, whereas we preferred to make no
further approximations beyond the modeling of the penetration factors
discussed in Sect.~\ref{sec:penetration}. Note that a comparison with
the work of \cite{Coc:2006sx} is not possible, since there any
variation of the fine-structure constant is tied to the variation of
certain Yukawa couplings.

All in all our results indicate that the BBN abundance for
$\nuc{7}{Li}+\nuc{7}{Be}$ is less sensitive and the abundance of
$\nuc{3}{H}+\nuc{3}{He}$ is more sensitive to variations of the value
of the electromagnetic fine-structure constant $\alpha$ than what was
determined earlier. Note that such a reduced sensitivity on nuclear
quantities, such as binding energies \textit{etc.} was also observed
in~\cite{Meissner:2022dhg}\,. There it was also found that this is
mainly due to inclusion of the temperature-dependent changes in the
rates. Unfortunately, the primordial abundance of
$\nuc{3}{H}+\nuc{3}{He}$ is not known precisely enough to lead to any
implications and the nominal prediction for the
$\nuc{7}{Li}+\nuc{7}{Be}$ abundance is too large anyway.

If we focus on the deuterium and $\nuc{4}{He}$ abundance ratios alone
we can extract from the observationally based data bounds on the
variation $\delta_\alpha$ of the value of the fine-structure constant
as listed in Table~\ref{tab:alphavar} for the four programs
considered here,
\begin{table}[!htb]
  \caption{%
    Lower ($\delta_\alpha^{\textnormal{min}}$) and upper
    ($\delta_\alpha^{\textnormal{max}}$) limits for the variation
    $\delta_\alpha$ of the fine-structure constant $\alpha=
    \alpha_0\,(1+\delta_\alpha)$ determined such that the resulting
    abundance lies within the error bounds of the observationally based
    abundance ratios for $\nuc{2}{H}$ and $\nuc{4}{He}$ given
    in~\cite{Workman:2022ynf}.
    \label{tab:alphavar}
  }
  \begin{ruledtabular}
    \begin{tabular}[c]{@{}l%
      @{\extracolsep{1.5em}}d{3.4}%
      @{\extracolsep{1.5em}}d{3.4}%
      @{\extracolsep{1.5em}}d{3.4}%
      @{\extracolsep{1.5em}}d{3.4}%
      @{}}
      \texttt{code}
      & \multicolumn{2}{c}{$\nuc{2}{H}$}
      & \multicolumn{2}{c}{$Y_p$}
      \\
      & \mc{$\delta_\alpha^{\textnormal{min}}$}
      & \mc{$\delta_\alpha^{\textnormal{max}}$}
      & \mc{$\delta_\alpha^{\textnormal{min}}$}
      & \mc{$\delta_\alpha^{\textnormal{max}}$}
      \\
      \colrule
      \texttt{NUC123}     &  0.002 &  0.008 & -0.011 & 0.006 \\
      \texttt{PArthENoPE} & -0.005 &  0.000 & -0.014 & 0.003 \\
      \texttt{AlterBBN}   & -0.007 & -0.001 & -0.018 & 0.000 \\
      \texttt{PRIMAT}     & -0.004 &  0.001 & -0.015 & 0.003 \\
    \end{tabular}
  \end{ruledtabular}
\end{table}
showing that one can allow for a variation of the fine-structure
constant $\alpha$ by less than 2\% on the basis of the results
obtained with all programs considered here using the current value
for the baryon-to-photon ratio $\eta=6.14\cdot 10^{-10}$, given
in~\cite{Workman:2022ynf}\,.  The values for the $\nuc{4}{He}$ mass
ratio $Y_p$ obtained with all four programs are rather consistent
and 
the range $[-0.018,0.006]$ which is more restrictive than the rough
estimate $|\delta_\alpha|<0.1$ quoted in~\cite{Nollett:2002da,
Bergstrom:1999wm} and the limit $|\delta_\alpha| \le 0.019$ mentioned
in~\cite{Dent:2007zu}.  The values found on the basis of the
deuterium number ratio show a larger spread, mainly because the
nominal values, see Table~\ref{tab:finab}, vary more strongly for the
four programs. In spite of this we can determine the range
$[-0.007,0.008]$\,, also still more restrictive than the (1$\sigma$)
range $[-0.04,0.10]$ of~\cite{Nollett:2002da}. Our new restrictions
on the variation of $\alpha$ are also stronger than found earlier in
the NLEFT analysis of the triple-alpha process in hot, old
stars~\cite{Epelbaum:2013wla,Lahde:2019yvr}.
 
From a comparison of Tables~\ref{tab:rmebeta}-\ref{tab:rmenuclrat} we
also see that the linear response for $Y_p$ due to variations in the
$\beta$-decay is the dominant effect. Indeed, as argued
in~\cite{Bergstrom:1999wm}, the variation of $Y_p$ with $\alpha$
mainly depends on the variation of the proton-neutron mass difference
with $\alpha$\,, \textit{i.e.} on $\delta m_{\textnormal{\tiny{QED}}}$
that enters the $n \to p$ weak decay.

As was done previously in Ref.~\cite{Nollett:2002da} we also studied
to what extend the results presently obtained vary with variations of
the baryon-to-photon ratio $\eta$ and found that our results for the
linear response coefficients $c_1$ do not change significantly if
$\eta$ is varied within the error range quoted
in~\cite{Workman:2022ynf}, $\eta_{10} = \eta\cdot 10^{10} = 6.143 \pm
0.190$\,. With the values of the primordial abundance ratios for $d,
\nuc{4}{He}$ and $\nuc{7}{Li}+\nuc{7}{Be}$ mentioned in
PDG~\cite{Workman:2022ynf}
\begin{figure}[!htb]
  \centering
  \includegraphics[width=1.0\columnwidth]{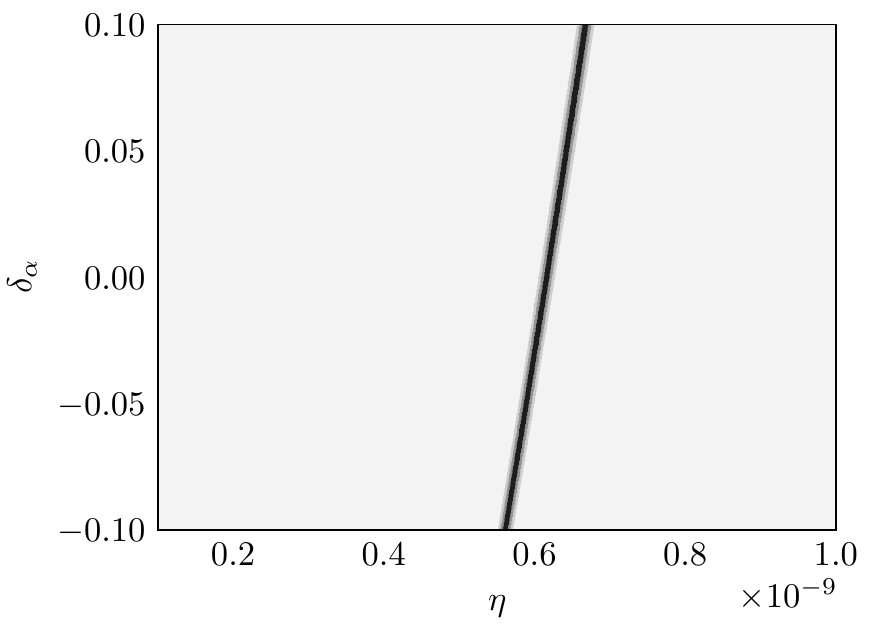}
  \caption{\label{fig:D_constr}%
    Restriction on the parameters $\delta_\alpha$ and $\eta$ based on
    the experimental value of $Y_{\nuc{2}{H}}/Y_H$ from
    ~\cite{Workman:2022ynf}\,. Shown are the corresponding $1\sigma$ (black),
    $2\sigma$ (dark gray) and $3\sigma$ (light gray) regions.
  }
\end{figure}
\begin{figure}[!htb]
  \centering
  \includegraphics[width=1.0\columnwidth]{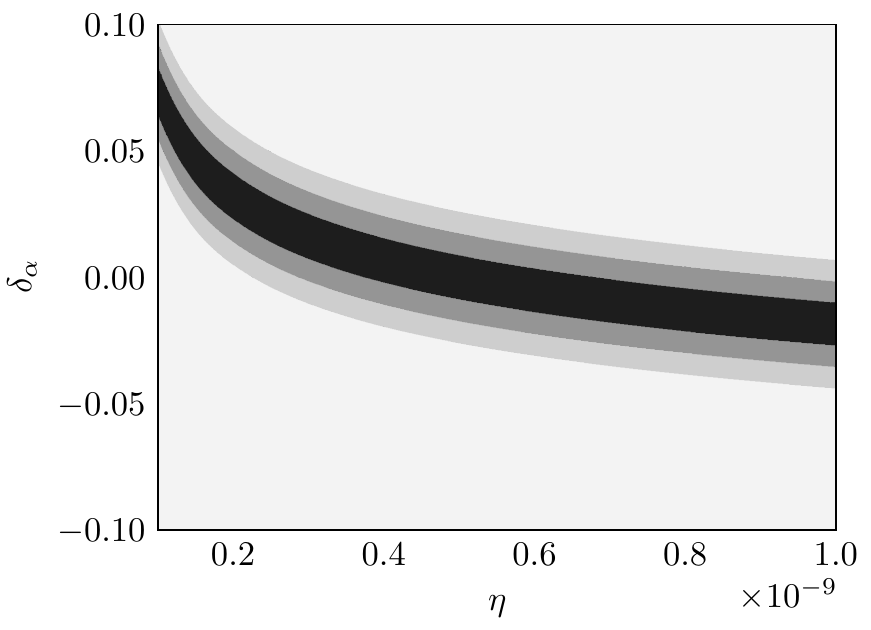}
  \caption{\label{fig:He_constr}%
    Restriction on the parameters $\delta_\alpha$ and $\eta$ based on
    the experimental value of $Y_p$ from
    ~\cite{Workman:2022ynf}\,. Shown are the corresponding $1\sigma$ (black),
    $2\sigma$ (dark gray) and $3\sigma$ (light gray) regions.
  }
\end{figure}
\begin{figure}[!htb]
  \centering
  \includegraphics[width=1.0\columnwidth]{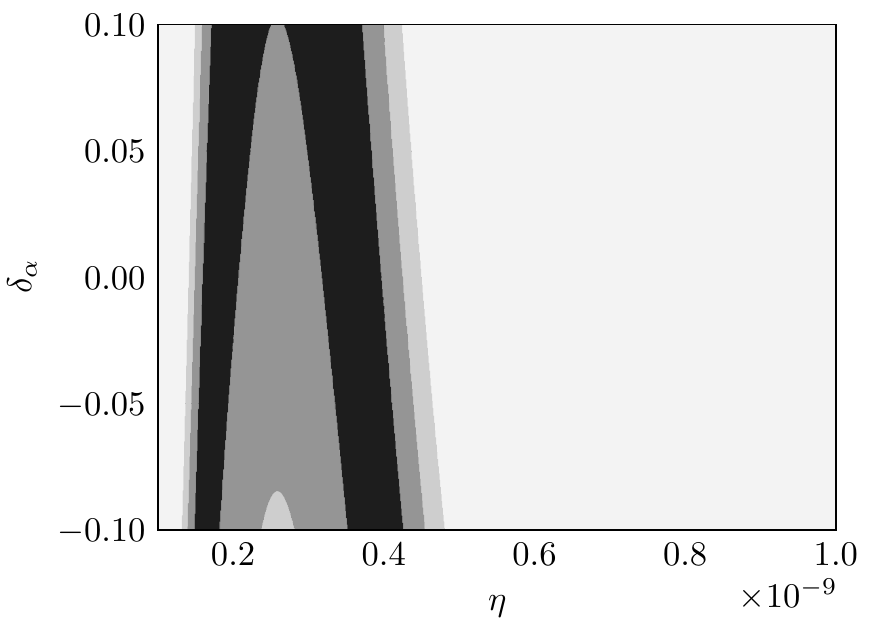}
  \caption{\label{fig:Li_constr}%
    Restriction on the parameters $\delta_\alpha$ and $\eta$ based on
    the experimental value of $Y_{(\nuc{7}{Li}+\nuc{7}{Be})}/Y_H$ from
    ~\cite{Workman:2022ynf}\,. Shown are the corresponding $1\sigma$ (black),
    $2\sigma$ (dark gray) and $3\sigma$ (light gray) regions.
  }
\end{figure}
we can derive parameter ranges for restricting $\delta_\alpha$ and
$\eta$ as presented in
Figs.~\ref{fig:D_constr}--\ref{fig:Li_constr}\,.  Note that we here
allowed for a variation of $\eta$ well beyond the currently accepted
limits quoted in~\cite{Workman:2022ynf}\,.  The results are similar to
those obtained in Ref.~\cite{Nollett:2002da} although the regions of
possible values for the $\delta_\alpha$- and $\eta$-values are
narrower here due to the newer, more precise observational data quoted
in~\cite{Workman:2022ynf}\,. The comparison of these results again
show that the value of the Li/Be abundance is incompatible with the
other data and we therefore refrain from any conclusions concerning
possible variations of $\alpha$ on the basis of the $\nuc{7}{Li}$
observation.

\section{\label{sec:summary}Summary} 

In the present paper we investigated the impact of variations in the
value of the fine-structure constant $\alpha$ on the abundances of the
light elements, \textit{viz.} $\nuc{2}{H}, \nuc{3}{H}+\nuc{3}{He},
\nuc{4}{He}, \nuc{6}{Li}$ and $\nuc{7}{Li}+\nuc{7}{B}$ in primordial
nucleosynthesis (BBN), keeping all other fundamental parameters fixed
on their values obtained in our universe. In order to estimate
possible model dependences concerning \textit{e.g.} the number of
reactions in the BBN nuclear network, the parameterizations of the
nuclear rates or the manner in which the corresponding rate equations
are numerically solved, we compared the results obtained by using four
different publicly available codes. Ideally such an investigation
requires an accurate \textit{ab initio} theory of nuclear reactions
accounting for all possible electromagnetic effects. Unfortunately,
however, for reactions involving the strong nuclear interaction this
is only realized for the leading nuclear reaction in the BBN network,
the $n + p \to d + \gamma$ reaction in the framework of pionless
EFT. For all other reactions
of this kind we rely on modifications of experimentally
determined reaction cross sections, trying to account for
electromagnetic effects, such as penetration factors, modeling the
suppression due to the Coulomb barrier in channels involving charged
particles as well as changes in the binding energies of nuclides due
to the Coulomb repulsion of the protons and hence the $Q$-values of
the nuclear reactions where these are involved. To this end we made
new parameterizations of the cross sections of the 18 leading nuclear
reactions in the BBN network using current experimental data compiled
by EXFOR.  We made an assumption about the $\alpha$ dependence of the
penetration factors which differs from the Gamow-factor form that was
used in previous investigations and used novel estimates for the
Coulomb contribution to nuclear binding energies based on a recent
\textit{ab initio} calculation in the framework of NLEFT in order to
determine the $\alpha$-dependence of the nuclear binding energies and
the corresponding $Q$-values. A further new ingredient for studying
the $\alpha$-dependence of the weak $\beta$-decays in the BBN network
is a novel value for the electromagnetic contribution to the
neutron-proton mass difference, which is slightly smaller than what
has been used before. All these new inputs were then used to determine
the variation of the reaction rates with varying $\alpha$. Here, we
found in particular that the variation of the reaction rates depends
on the temperature, a feature that seems to have been ignored in
previous investigations. We found consistent results with all four
codes mentioned above and hence conclude that the model-dependence
concerning the specific treatment of the BBN network is of minor
importance for the $\alpha$-dependence of the primordial abundances
studied here. The results for the linear response do, however, deviate
significantly from older results, in particular for the
$\alpha$-dependence of the abundances of $\nuc{3}{H}+\nuc{3}{He}$ and
$\nuc{7}{Li}+\nuc{7}{Be}$, the former being much larger and the
latter much smaller than found previously. Unfortunately, in the
standard Big Bang scenario used here, the nominal abundance ratio
$Y_{\nuc{7}{Li}+\nuc{7}{Be}}/Y_{\nuc{}H}$ exceeds the current
observationally based determination by a factor of three, a feature
known as the lithium-problem that is not solved in the present
treatment. This then also impedes a determination of consistent bounds
on the value of the fine-structure constant from all available
primordial abundance data. Using the observations for $\nuc{2}{H}$ and
$\nuc{4}{He}$ alone, we can nevertheless state that these data would
limit a possible variation of $\alpha$ to $|\delta_\alpha| <
0.02$. This is a stronger bound than found earlier in comparable
investigations.

An investigation of the kind presented here heavily relies on the
modeling of electromagnetic effects in the cross section data (or,
equivalently the astrophysical $S$-factors) of the relevant nuclear
reactions in the BBN-network. Here we opted for a specific form of
Coulomb penetration factors that differ from Gamow-factors used before
and stressed the relevance of the temperature dependence of the
variation of $\alpha$ in the reaction rates that resulted from
numerically integrating
$
\gamma(\alpha;T) \propto \intdif{0}{\infty}{E}\,E\,\sigma(\alpha;E)\,\exp(E/kT)
$.
It seems that further progress with the purpose of using primordial
nucleosynthesis as a laboratory for exploring our understanding of
fundamental physics, apart from astrophysical or cosmological aspects
will be feasible only if \textit{ab initio} theories describing the
relevant nuclear reactions including electromagnetic effects become
available. NLEFT appears to be a promising framework for doing just
that, see {\it e.g.}
Refs.~\cite{Elhatisari:2015iga,Elhatisari:2021eyg}.


\begin{acknowledgments}
  We are grateful to Serdar Elhatisari for providing the expectation
  values of the electromagnetic contribution to the binding energies of
  the nuclei considered here.  Furthermore we thank J\"urg Gasser,
  Heiri Leutwyler and Andreas Nogga for pointing out relevant
  references.  This project is part of the ERC Advanced Grant ``EXOTIC''
  supported the European Research Council (ERC) under the European
  Union's Horizon 2020 research and innovation programme (grant
  agreement No. 101018170), We further acknowledge support by the
  Deutsche Forschungsgemeinschaft (DFG, German Research Foundation) and
  the NSFC through the funds provided to the Sino-German Collaborative
  Research Center TRR110 ``Symmetries and the Emergence of Structure in
  QCD'' (DFG Project ID 196253076 - TRR 110, NSFC Grant
  No. 12070131001), the Chinese Academy of Sciences (CAS) President's
  International Fellowship Initiative (PIFI) (Grant No. 2018DM0034) and
  Volkswagen Stiftung (Grant No. 93562).
\end{acknowledgments}

\appendix

\section{\label{app:params}Parameterizations for $S$-factors and cross sections}

For the relevant reactions treated here almost all $S$-factors,
related to the cross section $\sigma$ as
\begin{equation}
  \label{eq:Ssigma}
  S(E) = E\,\sigma(E)\,\expo{\sqrt{{E_G^i}/{E}}}
\end{equation}
with $E_G^i$ given by Eq.~\eqref{eq:GamowEi}, 
can be written as
\begin{equation}
  \label{eq:SR}
  S(E) = S_0\,R(E;a_1,a_2,a_3,q_1,q_2,q_3)~,
\end{equation}
with $S_0$ in units of MeV\,mb, and where
\begin{eqnarray}
  \label{eq:SRatPol}
  \lefteqn{
  R(E;a_1,a_2,a_3,q_1,q_2,q_3)
  }
  \nonumber\\
  &&\qquad:=
     \frac{1 + a_1\,E + a_2\,E^2 + a_3\,E^3}{1 + q_1\,E + q_2\,E^2 + q_3\,E^3}~,
\end{eqnarray}
is a rational function of the center-of-mass (CMS) kinetic energy $E$
(given in MeV).  There are, however, some reactions where resonances
occur in the energy range considered here. For these, we can
parameterize the $S$-factor as the rational function of
Eq.~\eqref{eq:SRatPol} combined with relativistic Breit-Wigner
functions. The parameters for the relativistic Breit-Wigner functions
of the form
\begin{equation}\label{eq:BWfunc}
  BW(E ; b, \Gamma, M)
  =
  \frac{b}{\Gamma^2 M^2 + \left(E^2 - M^2\right)^2}
\end{equation}
can be found in Table~\ref{tab:Par_BW}, where $E, \Gamma$ and $M$ are
given in MeV.  The use of a non-relativistic Breit-Wigner function of
the form
\begin{equation}
  \label{eq:bwfunc}
  bw(E; b, \kappa, M)
  =
  \frac{b}{1+\kappa\,(E-M)^2}
\end{equation}
was found to be more appropriate for the reactions $\nuc{7}{Li} + d
\to n + \nuc{4}{He} + \nuc{4}{He}$ and $\nuc{7}{Be} + n \to
\nuc{7}{Li} + p$, the corresponding parameters ($\kappa$ in
\si{\mega\electronvolt^{-2}}) can also be found in
Table~\ref{tab:Par_BW}.

\begin{figure}[!htp]
  \centering
  \resizebox{1.0\columnwidth}{!}{\input{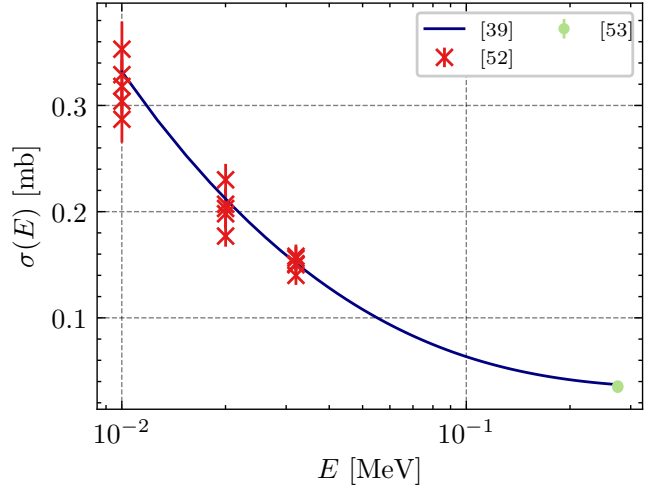}}
  \caption{\label{fig:npdg-cs}
    Calculation of the cross section for the $n + p \to d + \gamma$ reaction by
    \cite{Rupak:1999rk} compared to experimental data compiled by
    \cite{Exfor_web}.
  }
\end{figure}

\subsection{The $n + p \to d + \gamma$ reaction}

The cross section for the leading nuclear reaction of BBN, namely $n +
p \to d + \gamma$, was calculated according to the formulas\,,
\textit{viz.} Eqs.~(3.3)-(3.16) given in \cite{Rupak:1999rk} with the
parameters quoted there, also see Sect.~\ref{sec:npdg}. In
Fig.~\ref{fig:npdg-cs} this description is compared to the existing
data as compiled in~\cite{Exfor_web}\,.

\subsection{Other radiative capture reactions}\label{as:RadCap}

The parameters found by a fit of the parameters in
Eqs.~(\ref{eq:SR},\ref{eq:SRatPol}) to the data are displayed in
Table~\ref{tab:Par_RadCap} for most radiative capture reactions
treated here.

The parameterizations are compared to experimental data compiled by
EXFOR~\cite{Exfor_web} in Figs.~\ref{fig:dp3Heg}--\ref{fig:p6Li7Beg}.

\begin{table*}[!htp]
  \caption{%
    \label{tab:Par_RadCap}
    Fit parameters of the $S$-factor, see~Eq.(\ref{eq:Ssigma}), according to Eqs.~(\ref{eq:SR},\ref{eq:SRatPol}) and
    Eq.~(\ref{dagLi6}) for radiative capture reactions.
    $S_0$ is given in MeV\,mb; $a_k$ and $q_k$ in units of MeV${}^{-k}$\,.
  }
  \begin{ruledtabular}
    \begin{tabular}{@{}c%
      @{\extracolsep{4ex}}d{3.4}%
      @{\extracolsep{-0.8ex}}l%
      @{\extracolsep{4ex}}d{3.4}%
      @{\extracolsep{4ex}}d{3.4}%
      @{\extracolsep{4ex}}d{3.4}%
      @{\extracolsep{4ex}}d{3.4}%
      @{\extracolsep{4ex}}d{3.4}%
      @{\extracolsep{4ex}}d{3.4}%
      @{}}
      {Reaction} & \multicolumn{2}{c}{$S_0$} & \mc{$a_1$} & \mc{$a_2$} & \mc{$a_3$} & \mc{$q_1$} & \mc{$q_2$} & \mc{$q_3$}\\
      \colrule
      {$d + p \to \He{3} + \gamma$}           & 2.066 & $\times 10^{-4}$ & 30.431 & 14.943 &   0     & -0.032 & 0.035 & 0\\
      {$d + \He{4} \to \Li{6} + \gamma$}      & 3.162 & $\times 10^{-6}$ & -3.163 & 15.271 &  -0.633 &  0     & 0     & 0\\
      {$\T + p \to \He{4} + \gamma$}          & 1.875 & $\times 10^{-3}$ & 10.773 & 32.613 & 113.836 &  0     & 0     & 8.919 \times 10^{-3}\\      
      {$\T + \He{4} \to \Li{7} + \gamma$}     & 1.057 & $\times 10^{-1}$ & -1.378 &  1.106 &  0      &  0.128 & 0     & 0 \\
      {$\He{3} + \He{4} \to \Be{7} + \gamma$} & 4.912 & $\times 10^{-1}$ & -0.908 &  0.336 &  0      & -0.610 & 0.247 & 0 \\
      {$\Li{6} + p \to \Be{7} + \gamma$}      & 5.000 & $\times 10^{-2}$ &-13.863 & 53.532 & 14.977  &-10.907 & 33.652& 0 \\
    \end{tabular}
  \end{ruledtabular}
\end{table*} 

\begin{figure}[!htp]
  \centering 
    \resizebox{1.0\columnwidth}{!}{\input{dpgHe3_log.pgf}}
    \caption{\label{fig:dp3Heg}
      Fit (red curve, color online) of the $S$-factor for the
      $d + p \to \He{3} + \gamma$
      reaction compared to data  compiled by EXFOR \cite{Exfor_web}\,.
    }
\end{figure}

\begin{figure}[!htp]
  \centering
  \resizebox{1.0\columnwidth}{!}{\input{H3pga_log.pgf}}
  \caption{\label{fig:3Hp4Heg}
    Fit (red curve, color online) of the $S$-factor for the
    $\T + p \to \He{4} + \gamma$
    reaction compared to data compiled by EXFOR \cite{Exfor_web}\,.
  }
\end{figure}

\begin{figure}[!htp]
  \centering
  \resizebox{1.0\columnwidth}{!}{\input{H3agLi7_log.pgf}}
  \caption{\label{fig:3H4He7Lig}
    Fit (red curve, color online) of the $S$-factor for the
    $\T + \He{4} \to \Li{7} + \gamma$
    reaction compared to data compiled by EXFOR \cite{Exfor_web}\,.
  }
\end{figure}

\begin{figure}[!htp]
  \centering
  \resizebox{1.0\columnwidth}{!}{\input{He3alphagBe7_log.pgf}}
  \caption{\label{fig:4He4He7Beg}
    Fit (red curve, color online) of the $S$-factor for the
    $\He{3} + \He{4} \to \Be{7} + \gamma$
    reaction compared to data compiled by EXFOR \cite{Exfor_web}\,.
  }
\end{figure}

\begin{figure}[!htp]
  \centering
    \resizebox{1.0\columnwidth}{!}{\input{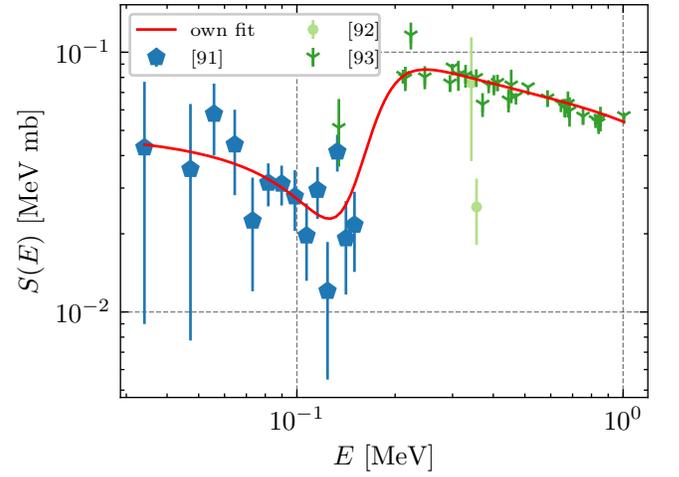}}
    \caption{\label{fig:p6Li7Beg}
      Fit (red curve, color online) of the $S$-factor for the
      $\Li{6} + p \to \Be{7} + \gamma$
      reaction compared to data compiled by EXFOR \cite{Exfor_web}\,.
    }
\end{figure}

The only exception is the reaction ${d + \He{4} \to \Li{6} + \gamma}$,
where a resonance appears. In this case the $S$-factor is given by the
sum of a cubic polynomial in $E$ and a relativistic Breit-Wigner
function:

\begin{eqnarray}
  \label{dagLi6}
  S(E)
  &=&
      S_0\,\left(
      1 + a_1 E + a_2 E^2 + a_3 E^3\right)
  \nonumber\\
  && 
      \qquad +
     BW(E ; b, \Gamma, M)
\end{eqnarray}
with the parameters listed in Tables~\ref{tab:Par_RadCap} and
\ref{tab:Par_BW}.  This parameterization is compared to experimental
data compiled by EXFOR~\cite{Exfor_web} in Fig.~\ref{fig:d4He6Lig}\,.

\begin{figure}[!htp]
  \centering
    \resizebox{1.0\columnwidth}{!}{\input{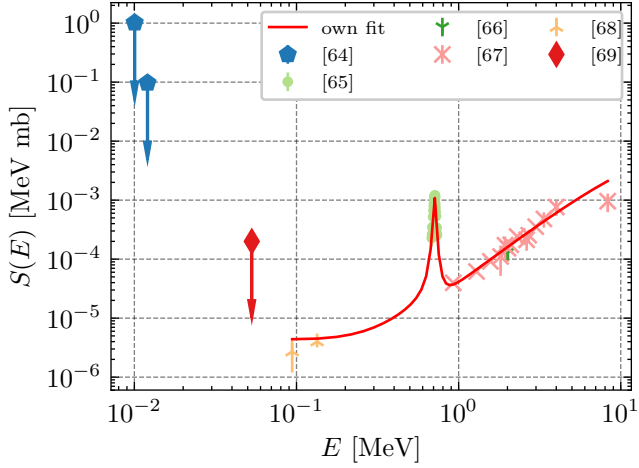}}
    \caption{\label{fig:d4He6Lig}
      Fit (red curve, color online) of the $S$-factor for the
      $d + \He{4} \to \Li{6} + \gamma$
      reaction compared to data compiled by EXFOR \cite{Exfor_web}\,.
      The three upper limits for $E<0.1~\textnormal{MeV}$, denoted by arrows,
      were not included in the fit.
    }
\end{figure}

\subsection{Charged particle reactions}

As in \ref{as:RadCap}, we fitted the $S$-factors according to
Eqs.~(\ref{eq:SR},\ref{eq:SRatPol}). The parameters are displayed in
Table~\ref{tab:Par_Charged} for most charged particle reactions
treated here.

\begin{table*}[!htp]
  \caption{%
    \label{tab:Par_Charged}
    Fit parameters of the $S$-factor, see~Eq.(\ref{eq:Ssigma}), according
    to Eqs.\,(\ref{eq:SR},\ref{eq:SRatPol},\ref{Li6p},\ref{Li7p}) and (\ref{Li7d})
    for charged particle reactions.
    $S_0$ is given in MeV\,mb; $a_k$ and $q_k$ in units of MeV${}^{-k}$.
    For these reactions $q_3 = 0$\,.
  }
  \begin{ruledtabular}
    \begin{tabular}{@{}l%
      @{\extracolsep{2.5em}}c%
      @{\extracolsep{3.5em}}d{5.3}%
      @{\extracolsep{3em}}d{5.3}%
      @{\extracolsep{1.2em}}d{3.3}%
      @{\extracolsep{1.2em}}d{3.3}%
      @{\extracolsep{3em}}d{2.3}%
      @{\extracolsep{1.2em}}d{2.3}%
      @{}}
      \mc{Reaction} & \mc{Energy range}     & \mc{$S_0$} & \mc{$a_1$} & \mc{$a_2$} & \mc{$a_3$} & \mc{$q_1$} & \mc{$q_2$} \\
      \colrule
      {${d + d \to \He{3} + n}$}            && 54.908  &     6.942 &   0.378 &  0     &  0.636 &  -0.018 \\  
      {${d + d \to p + \T}$}                && 70.667  &    27.281 & 136.744 &  0     & 38.369 &   9.531 \\
      \colrule
      \multirow{2}{*}{$\T + d \to n + \He{4}$} & {$E < \SI[round-precision = 2]{0.28}{\mega\electronvolt}$}
                                            & 10800.846 &  -1.974 & 18.252 &  0     & -24.464 & 244.175 \\
                                               & {$E \geq \SI[round-precision = 2]{0.28}{\mega\electronvolt}$}
                                            & -2116.168 &   0.137 &  0.527 & -0.038 &  -8.747 & 0 \\
      \colrule
      \multirow{2}{*}{$\He{3} + d \to p + \He{4}$} & {$E < \SI[round-precision = 2]{0.25}{\mega\electronvolt}$}
                                            &  6703.216 &  -8.823 & 27.654 & -2.772 &  -9.380 & 24.921\\
                                                   &  {$E \geq \SI[round-precision = 2]{0.25}{\mega\electronvolt}$}
                                            & 10663.275 &  -0.899 &  1.562 & -0.033 & -6.664 & 20.204 \\
      \colrule
      {${\Li{6} + p \to \He{3} + \He{4}}$}  &&  288.587 &  -0.305 & -1.494 &  0.981 &  0     &  0 \\
      \colrule 
      \multirow{2}{*}{{$\Li{7} + p \to \He{4} + \He{4}$}} & {$E \leq \SI[round-precision = 1]{4.1}{\mega\electronvolt}$}
                                            &   338.062 &   0.731 & -0.102 &  0     &  0     &  0 \\
                    & {$E > \SI[round-precision = 1]{4.1}{\mega\electronvolt}$}
                                            & 12312.399 &  -0.472 &  0.057 &  0     &  0     &  0 \\
    \colrule
      {$\Be{7} + d \to p + \He{4} + \He{4}$} && 684.412 &  -0.554 &  0.142 &  0     & -0.535 &  0.077 \\   
      {$\Li{7} + d \to n + \He{4} + \He{4}$} &&  2968.470 &   8.279 & -0.308 &  0     & 54.611 &  0 \\
    \end{tabular} 
  \end{ruledtabular}
\end{table*}

For the reaction ${\Li{6} + p \to \He{3} + \He{4}}$ which has two
resonances, the $S$-factor is given by an expression of the form
\begin{eqnarray}
  \label{Li6p}
  S(E)
  &=&
  S_0\,\left(1 + a_1 E + a_2 E^2 + a_3 E^3\right)
     \nonumber\\
  &&
     \hspace*{-2em}
     \times
     BW(E ; b_1, \Gamma_1, M_1)
     \times
     BW(E ; b_2, \Gamma_2, M_2)\,\,
\end{eqnarray}
  and for the reaction $\Li{7} + p \to \He{4} + \He{4}$
  with one resonance the $S$-factor
  is given by
\begin{eqnarray}
  \label{Li7p}
  S(E)
  &=&
  S_0\,\left(1 + a_1 E + a_2 E^2 + a_3 E^3\right)
     \nonumber\\
  &&
     \,\,\,
     \times
     BW(E ; b_1, \Gamma_1, M_1)~.
\end{eqnarray}

For the reaction ${\Li{7} + d \to n + \He{4} + \He{4}}$ a
parameterization of the form
\begin{eqnarray}
  \label{Li7d}
  S(E)
  &=&
      S_0\,\frac{1 + a_1\,E + a_2\,E^2}{1 + q_1\,E}
     \,\,\,+\,\,
     bw(E ; b_1, \kappa_1, M_1)
     \nonumber\\
  &&
     \qquad\,\,\,+\,\,
     bw(E ; b_2, \kappa_2, M_2)
\end{eqnarray}
was used. The parameters of the Breit-Wigner functions can be found in
Table~\ref{tab:Par_BW}\,.

These $S$-factor fits are compared to experimental data compiled by in
EXFOR~\cite{Exfor_web} in Figs.~\ref{fig:ddn3He}--\ref{fig:7Bedpaa}\,.

\begin{figure}[!htp]
  \centering
    \resizebox{1.0\columnwidth}{!}{\input{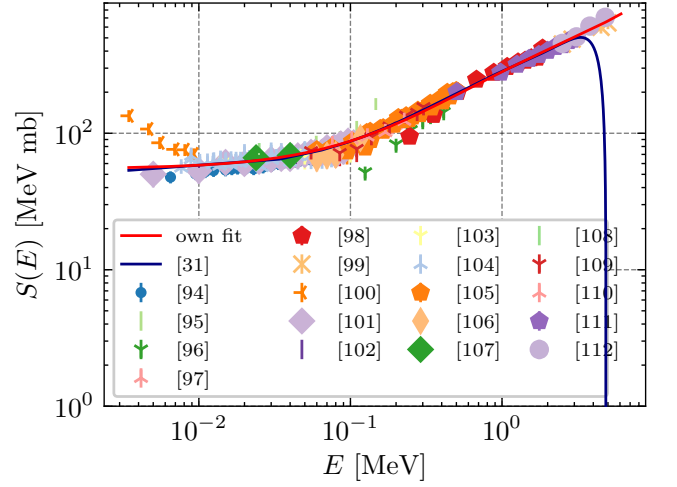}}
    \caption{\label{fig:ddn3He}
      Fit (red curve, color online) of the $S$-factor for the
      $d + d \to \He{3} + n$
      reaction compared to data compiled by EXFOR \cite{Exfor_web}\,.
    }
\end{figure}

\begin{figure}[!htp]
  \centering
    \resizebox{1.0\columnwidth}{!}{\input{ddpH3_log.pgf}}
    \caption{\label{fig:ddp3H}
      Fit (red curve, color online) of the $S$-factor for the
      $d + d \to p + \T$
      reaction compared to data compiled by EXFOR \cite{Exfor_web}\,.
    }
\end{figure}

\begin{figure}[!htp]
  \centering
    \resizebox{1.0\columnwidth}{!}{\input{H3dnHe4_log.pgf}}
    \caption{\label{fig:tdn4He}
      Fit (red curve, color online) of the $S$-factor for the
      $\T + d \to n + \He{4}$
      reaction compared to data compiled by EXFOR \cite{Exfor_web}\,.
    }
\end{figure}

\begin{figure}[!htp]
  \centering
    \resizebox{1.0\columnwidth}{!}{\input{He3dpHe4_log.pgf}}
    \caption{\label{fig:3Hedp4He}
      Fit (red curve, color online) of the $S$-factor for the
      $\He{3} + d \to p + \He{4}$
      reaction compared to data compiled by EXFOR \cite{Exfor_web}\,.
    }
\end{figure}

\begin{figure}[!htp]
  \centering
    \resizebox{1.0\columnwidth}{!}{\input{Li6paHe3_log.pgf}}
    \caption{\label{fig:6Lipa3He}
      Fit (red curve, color online) of the $S$-factor for the
      $\Li{6} + p \to \He{3} + \He{4}$
      reaction compared to data compiled by EXFOR \cite{Exfor_web}\,.
    }
\end{figure}

\begin{figure}[!htp]
  \centering
    \resizebox{1.0\columnwidth}{!}{\input{Li7paHe4_log.pgf}}
    \caption{\label{fig:7Lipa4He}
      Fit (red curve, color online) of the $S$-factor for the
      $\Li{7} + p \to \He{4} + \He{4}$
      reaction compared to data compiled by EXFOR \cite{Exfor_web}\,.
    }
\end{figure}

\begin{figure}[!htp]
  \centering
    \resizebox{1.0\columnwidth}{!}{\input{Li7dnBe8_log.pgf}}
    \caption{\label{fig:7Lidnaa}
      Fit (red curve, color online) of the $S$-factor for the
      $\Li{7} + d \to n + \He{4} + \He{4}$
      reaction compared to data compiled by EXFOR \cite{Exfor_web}\,.
    }
\end{figure}

\begin{figure}[!htp]
  \centering
    \resizebox{1.0\columnwidth}{!}{\input{Be7dpBe8_log.pgf}}
    \caption{\label{fig:7Bedpaa}
      Fit (red curve, color online) of the $S$-factor for the
      $\Be{7} + d \to p + \He{4} + \He{4}$
      reaction compared to data compiled by EXFOR \cite{Exfor_web}\,.
    }
\end{figure}

\subsection{Neutron-induced reactions}

For neutron capture reactions the cross section is written as 
\begin{equation}
  \label{eq:ncap}
  \sigma(E) = S^{[n]}(E) / \sqrt{E},
\end{equation}
implying that the function $S^{[n]}$ is given in units of 
$\si{(\mega\electronvolt)^{1/2} \milli\barn}$.
The $S$-factor then reads
\begin{equation}
  \label{eq:SSn}
  S(E) = \sigma(E)\,E = S^{[n]}(E)\,\sqrt{E}\,,
\end{equation}
since for neutron-induced reactions the Gamow-factor is unity.  For
neutron-induced reactions we give parameterizations of $S^{[n]}(E)$ in
terms of
\begin{equation}
  \label{eq:SnR}
   S^{[n]}_0\,R(E;a_1,a_2,a_3,q_1,q_2,q_3)
\end{equation}
with the rational function $R$ of Eq.~(\ref{eq:SRatPol}) and the
Breit-Wigner functions of Eqs.~(\ref{eq:BWfunc},\ref{eq:bwfunc}).
$S^{[n]}_0$ is then given in units of MeV${}^{1/2}$\,mb.

\begin{figure}[!htp]
  \centering
    \resizebox{1.0\columnwidth}{!}{\input{He3npH3_log.pgf}}
    \caption{\label{fig:3Henp3H}
      Fit (red curve, color online) of $S(E) = S^{[n]}(E)\,\sqrt{E}$ for the
      $\He{3} + n \to p + \T$
      reaction compared to data compiled by EXFOR \cite{Exfor_web}\,.
    }
\end{figure}

\begin{figure}[!htp]
  \centering
  \resizebox{1.0\columnwidth}{!}{\input{Be7npLi7_log.pgf}}
  \caption{\label{fig:7Benp7Li}
    Fit (red curve, color online) of $S(E) = S^{[n]}(E)\,\sqrt{E}$ for the
    $\Be{7} + n \to p + \Li{7}$
    reaction compared to data compiled by EXFOR \cite{Exfor_web}\,.
  }
\end{figure}

\begin{figure}[!htp]
  \centering
  \resizebox{1.0\columnwidth}{!}{\input{Be7naHe4_log.pgf}}
  \caption{\label{fig:7Benaa}
    Fit (red curve, color online) of $S(E) = S^{[n]}(E)\,\sqrt{E}$ for
    the $\Be{7} + n \to \He{4} + \He{4}$ reaction compared to data
    compiled by EXFOR \cite{Exfor_web}\,.
  }
\end{figure}

\begin{table*}[!htp]
  \caption{%
    \label{tab:Par_Neutron}
      Fit parameters of the function $S^{[n]}$,
      see~Eq.(\ref{eq:ncap}), according to
      Eqs.~(\ref{eq:SnR},\ref{eq:SRatPol}) neutron-induced reactions.
      $S^{[n]}_0$ is given in MeV${}^{1/2}$\,mb, and $a_k$ and $q_k$ in
      units of MeV${}^{-k}$\,.
    }
  \begin{ruledtabular}
    \begin{tabular}{@{}l%
      @{\extracolsep{3em}}c%
      @{\extracolsep{2.5em}}d{5.3}%
      @{\extracolsep{3em}}d{5.3}%
      @{\extracolsep{1em}}d{3.3}%
      @{\extracolsep{1em}}d{4.3}%
      @{\extracolsep{3em}}d{2.3}%
      @{\extracolsep{1em}}d{2.3}%
      @{\extracolsep{1em}}d{2.3}%
      @{}}
      \mc{Reaction} & \mc{Energy range} & \mc{$S^{[n]}_0$} & \mc{$a_1$} & \mc{$a_2$} & \mc{$a_3$} & \mc{$q_1$} & \mc{$q_2$} & \mc{$q_3$} \\
      \colrule
      \multirow{2}{*}{$\He{3} + n \to p + \T$} & {$E \leq \SI[round-precision = 2]{2.8}{\mega\electronvolt}$}
                                        &  715     &   20.814 &   0     &   6.8   & 38.681 & 27.876 & 12.637 \\
                    &  {$E > \SI[round-precision = 2]{2.8}{\mega\electronvolt}$}
                                        & 1691.556\footnote{here $S^{[n]}_0$ in units of MeV\,mb; to be divided by $\sqrt{E}$, $E$ in MeV. in order to yield $S^{[n]}$.}
                                                     &   -0.280 &   0.033 &  -0.001 & -0.234 &  0.024 &  0 \\
      \colrule
      \multirow{2}{*}{$\Be{7} + n \to \He{4} + \!\!\He{4}$} & {$E \leq \SI[round-precision = 2]{2.0}{\mega\electronvolt}$}
                                        &    0.381 &  22.875 &  7.931 &   0     &  0     &  0     & 0\\ 
                    &  {$E > \SI[round-precision = 2]{2.0}{\mega\electronvolt}$}
                                        &   -0.418 & -108.504 & 79.576 & -15.092 &  0     &  0     & 0\\ 
    \end{tabular}
  \end{ruledtabular}
\end{table*}

\begin{table*}[!htp]
  \caption{
    \label{tab:Par_BW}
    Fit parameters for resonances parameterized as Breit-Wigner
    functions in Eqs.~(\ref{eq:BWfunc},\ref{eq:bwfunc}).  $\Gamma_k$ and
    $M_k$ in MeV; $\kappa_k$ in MeV${}^{-2}$\,. The units for $b_k$ and
    $c$ depend on the context, see
    Eqs.~(\ref{dagLi6},\ref{Li6p},\ref{Li7p},\ref{Li7d}).
  }
  \begin{ruledtabular}
    \begin{tabular}{@{}l%
      @{\extracolsep{1.7em}}c%
      @{\extracolsep{1.7em}}d{4.3}%
      @{\extracolsep{1em}}d{2.3}%
      @{\extracolsep{1em}}d{1.3}%
      @{\extracolsep{2.7em}}d{3.3}%
      @{\extracolsep{1em}}d{5.3}%
      @{\extracolsep{1em}}d{1.3}%
      @{\extracolsep{2.7em}}d{3.3}%
      @{}}
      \mc{Reaction} & \mc{Energy range} & \mc{$b_1$} & \mc{$\Gamma_1$} & \mc{$M_1$} & \mc{$b_2$} & \mc{$\Gamma_2$} & \mc{$M_2$} & \mc{$c$} \\
      \colrule
      {$d + \He{4} \to \Li{6} + \gamma$} && \mc{$4.310 \times 10^{-7}$} &  0.028 & 0.711 &         &        &      &  \\
      {$\Li{6} + p \to \He{3} + \He{4}$} &&                   5113.917 &  0.654 & 1.187 & 104.696 & 13.972 & 8.72  &  0 \\
      \colrule
      \multirow{2}{*}{$\Li{7} + p \to \He{4} + \He{4}$} & {$E \leq \SI[round-precision = 1]{4.1}{\mega\electronvolt}$}
                                        &                       12.395 &  1.012 & 2.669  &        &        &      & \\
                    & {$E > \SI[round-precision = 1]{4.1}{\mega\electronvolt}$}
                                        &                       27.448 &  0.912 & 4.824  &        &        &      &  \\
      \colrule
      \multirow{2}{*}{$\Be{7} + n \to \He{4} + \He{4}$} & {$E \leq \SI[round-precision = 1]{2.0}{\mega\electronvolt}$}
                                        &                        4.023 &  0.825 & 0.887 &    2.035 & 0.455 & 3.482 &  0.156\\
                    & {$E > \SI[round-precision = 1]{2.0}{\mega\electronvolt}$}
                                        &                        9.331 &  1.246 & 1.346 &   39.447 & 0.723 & 3.114 & -0.023 \\ 
      \colrule
                    & & \mc{$b_1$} & \mc{$\kappa_1$} & \mc{$M_1$} & \mc{$b_2$} & \mc{$\kappa_2$} & \mc{$M_2$} & \mc{$c$} \\
      \colrule 
      {$\Li{7} + d \to n + \He{4} + \He{4}$} &&                  9820.6 &  82.387 & 0.6   &   8991.0 & 1963.84 & 0.8 & 0 \\ 
      {$\Be{7} + n \to \Li{7} + p$}          &&                  1.116 & 131.7   & 0.327 &   0     &    0    & 0   & 0 \\
    \end{tabular}
  \end{ruledtabular}
\end{table*}

Note that for the reaction ${\He{3} + n \to p + \T}$ for $E >
\SI{2.8}{\mega\electronvolt}$, the rational polynomial described by
the coefficients in Table~\ref{tab:Par_Neutron} still needs to be
divided by $\sqrt{E}$.  For this reaction the fit of the $S$-factor is
compared to experimental data compiled by EXFOR~\cite{Exfor_web} in
Fig.~\ref{fig:3Henp3H}\,.

For the reaction ${\Be{7} + n \to p + \Li{7}}$ the following
parameterization in terms of a non-relativistic Breit-Wigner function
and a polynomial in $\sqrt{E}$ was used:
\begin{eqnarray}
  \lefteqn{
  \label{eq:par7Benp7Li}
  S^{[n]}(E) = 1000.0
  }
  \nonumber\\
  &&
     \times\left\lbrace
      \begin{array}[c]{ll}
        bw(E; b, \kappa, M)&
        \\
        \quad+ 7.7874 - 47.778\,E^{\frac{1}{2}} &
        \\
        \quad+ 140.00\,E - 222.87\,E^{\frac{3}{2}} &
        \\
        \quad+ 201.84\,E^2 - 97.983\,E^{\frac{5}{2}} &
        \\
        \quad+ 19.773\,E^3 \,\,\si{(\mega\electronvolt)^{1/2} \milli\barn}\,, 
                                               & E \le 2.0\,\textnormal{MeV}\,,
        \\
        \SI[round-mode = places, round-precision = 3]{1139.6274903351064}{\,\,(\mega\electronvolt)^{1/2}\milli\barn}\,,
                                               & E >   2.0\,\textnormal{MeV}
      \end{array}
            \right.
            \nonumber\\
\end{eqnarray}
where again the CMS energy $E$ is given in MeV and the parameters of
the Breit-Wigner function can be found in Table~\ref{tab:Par_BW}. For
this reaction the $S$-factor fit is compared to experimental data
compiled by EXFOR~\cite{Exfor_web} in Fig.~\ref{fig:7Benp7Li}\,.

Finally, the form of the parameterization for the reaction
${\Be{7} + n \to \He{4} + \He{4}}$ reads
\begin{eqnarray}
  \label{eq:spcl}
  S^{[n]}(E)
  &=&
      S^{[n]}_0\, \left(1 + a_1\,E + a_2\,E^2 + a_3\,E^3 \right)
      \nonumber\\
  &\times&
           \bigl( c + 
           BW(E;b_1, \Gamma_1, M_1)
           +
           BW(E;b_2, \Gamma_2, M_2)
           \bigr)~.
           \nonumber
  \\
  &&
     \hspace*{10em}
\end{eqnarray} 
The $S$-factor is compared to experimental data compiled by EXFOR
\cite{Exfor_web} in Fig.~\ref{fig:7Benaa}\,.



\end{document}